\documentclass{article}

\usepackage{microtype}
\usepackage{graphicx}
\usepackage{subcaption}
\usepackage{booktabs} % for professional tables

\usepackage{hyperref}

\usepackage[accepted]{icml2026}

\usepackage{amsmath}
\usepackage{amssymb}
\usepackage{mathtools}
\usepackage{amsthm}

\usepackage{url}
\usepackage{float}
\usepackage{mdframed}
\usepackage{enumitem}
\usepackage{xcolor}
\usepackage{wrapfig}
\usepackage{forest}
\usepackage[most]{tcolorbox} %     
\usepackage{courier}
\usepackage[hypcap=true]{caption}
\usepackage{listings}
\usepackage{multirow}
\usepackage{makecell}

\usepackage[capitalize,noabbrev]{cleveref}

\theoremstyle{plain}

\theoremstyle{definition}

\theoremstyle{remark}

\definecolor{lightgreenbin}{RGB}{214,233,199}
\definecolor{lightbluebin}{RGB}{204,232,246}
\newcommand{\Sref}[1]{\S\ref{#1}}
\usepackage{soul}
\sethlcolor{yellow}
\soulregister\ref7

\usepackage[textsize=tiny]{todonotes}

\icmltitlerunning{AICrypto: Evaluating Cryptography Capabilities of Large Language Models}

\begin{document}

\twocolumn[
  \icmltitle{AICrypto: Evaluating Cryptography Capabilities of Large Language Models}



  \icmlsetsymbol{equal}{*}
  \icmlsetsymbol{cor}{$\dagger$}
  \icmlsetsymbol{iie}{1}
  \icmlsetsymbol{ucas}{2}
  \icmlsetsymbol{sqz}{3}
  \icmlsetsymbol{thu}{4}
  \icmlsetsymbol{xa}{5}
  \icmlsetsymbol{cai}{6}
  \begin{icmlauthorlist}
    \icmlauthor{Yu Wang}{equal,iie,ucas,sqz}
    \icmlauthor{Yijian Liu}{equal,iie,ucas}
    \icmlauthor{Liheng Ji}{equal,thu,sqz}
    \icmlauthor{Han Luo}{equal,thu}
    \icmlauthor{Wenjie Li}{equal,thu,sqz}
    \icmlauthor{Xiaofei Zhou}{iie,ucas,cor}
    \icmlauthor{Chiyun Feng}{iie,ucas}
    \icmlauthor{Puji Wang}{ucas}
    \icmlauthor{Yuhan Cao}{sqz}
    \icmlauthor{Geyuan Zhang}{iie,ucas}
    \icmlauthor{Xiaojian Li}{cai}
    \icmlauthor{Rongwu Xu}{thu,sqz}
    \icmlauthor{Yilei Chen}{thu,sqz,cor}
    \icmlauthor{Tianxing He}{thu,xa,sqz,cor}

  \end{icmlauthorlist}

  % \icmlaffiliation{iie}{Institute of Information Engineering, Chinese Academy of Sciences, Beijing, China}
  % \icmlaffiliation{ucas}{School of Cyber Security, University of Chinese Academy of Sciences, Beijing, China}
  % \icmlaffiliation{sqz}{Shanghai Qi Zhi Institute}
  % \icmlaffiliation{thu}{Institute for Interdisciplinary Information Sciences, Tsinghua University}
  % \icmlaffiliation{xa}{Xiongan AI Institute}
  % \icmlaffiliation{cai}{College of AI, Tsinghua University}

  \icmlcorrespondingauthor{Xiaofei Zhou}{zhouxiaofei@iie.ac.cn}
  \icmlcorrespondingauthor{Yilei Chen}{chenyilei@mail.tsinghua.edu.cn}
  \icmlcorrespondingauthor{Tianxing He}{hetianxing@mail.tsinghua.edu.cn}
  \icmlkeywords{Machine Learning, ICML}

  \vskip 0.2in
]



\printAffiliationsAndNotice{$^{*}$Equal contribution. $^1$Institute of Information Engineering, Chinese Academy of Sciences, Beijing, China $^2$School of Cyber Security, University of Chinese Academy of Sciences, Beijing, China $^3$Shanghai Qi Zhi Institute $^4$Institute for Interdisciplinary Information Sciences, Tsinghua University $^5$Xiongan AI Institute $^6$College of AI, Tsinghua University}  % no special notice (required even if empty)


\begin{abstract}

We build \textbf{AICrypto}, a comprehensive benchmark designed to evaluate the cryptography capabilities of large language models (LLMs). The benchmark comprises 135 multiple-choice questions, 150 capture-the-flag challenges, and 30 proof problems, covering a broad range of skills from knowledge memorization to vulnerability exploitation and formal reasoning. All tasks are carefully reviewed or constructed by cryptography experts to improve correctness and rigor. For each proof problem, we provide detailed scoring rubrics and reference solutions that enable automated grading, achieving high correlation with human expert evaluations. We introduce strong human expert performance baselines for comparison across all task types. Our evaluation of 17 leading LLMs reveals that state-of-the-art models match or even surpass human experts in memorizing cryptographic concepts, exploiting common vulnerabilities, and routine proofs. However, our analysis reveals that they still lack a deep understanding of abstract mathematical concepts and struggle with tasks that require multi-step reasoning and dynamic analysis. We hope this work could provide insights for future research on LLMs in cryptographic applications. Our code and dataset are available at \href{https://github.com/wangyu-ovo/aicrypto-agent}{https://github.com/wangyu-ovo/aicrypto-agent}.
\end{abstract}

\section{Introduction}

Modern cryptography is a complex, interdisciplinary field that forms the foundation of cybersecurity. It plays a vital role in everyday communication to military operations  \citep{rivest1978method,shamir1979share,NISTPQC}. With large language models (LLMs) gaining substantial mathematical and coding prowess~\citep{guo2025deepseek, o3o4mini}. This development raises an important question: \textit{What is the current state of LLMs' cryptographic capability?} 

Several studies evaluate the performance of LLMs on cybersecurity tasks \citep{shao2024nyuctfbench, zhang2025cybench, cvebench, zhang2025bountybenchdollarimpactai}, with some also touching on cryptographic scenarios. \citet{li2025cipherbank} assesses reasoning ability by examining how LLMs perform on decryption tasks. However, to the best of our knowledge, no comprehensive, cryptography-specific benchmark for LLMs exists. This gap stems from the inherent complexity and interdisciplinary nature of the field. First, cryptography spans both theoretical foundations and practical implementations. Second, modern cryptographic tasks often involve heavy, large-number computation, which LLMs are not good at.

\begin{figure*}[t]
  \centering
  \includegraphics[width=0.83\linewidth]{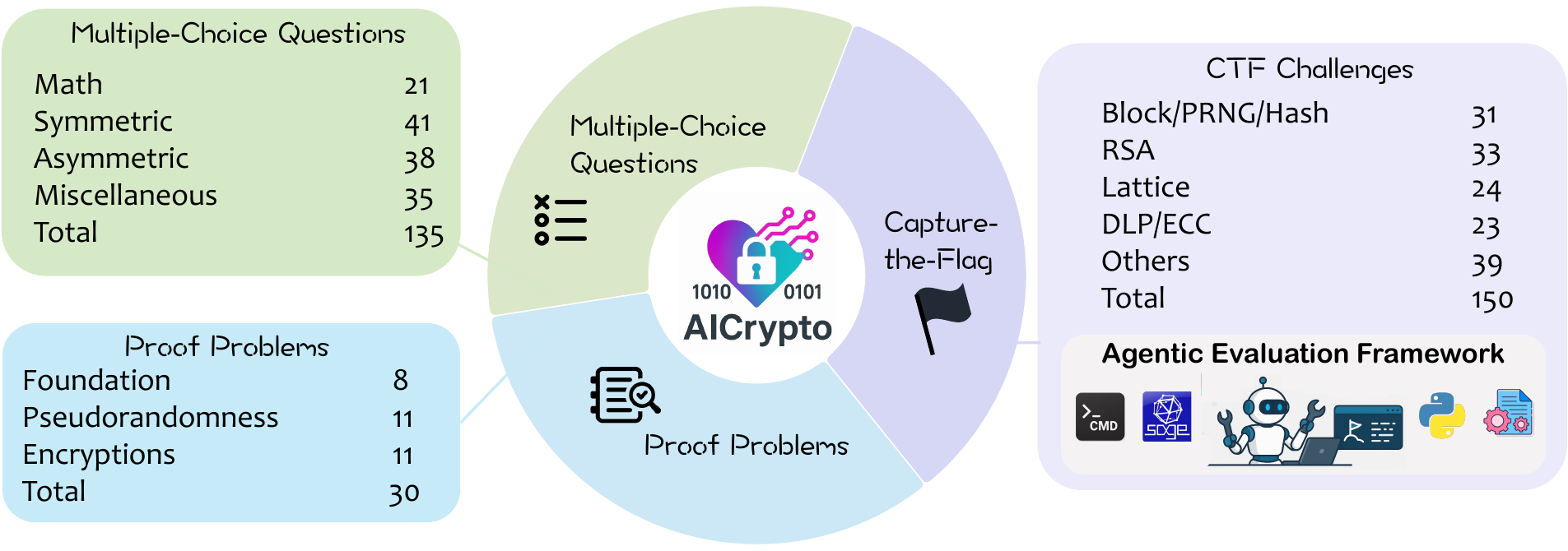}
  \caption{Overview of the AICrypto benchmark.}
  \vspace{-4mm}
  \label{fig:overview}
\end{figure*}

We present \textbf{AICrypto}, a comprehensive benchmark developed in extensive collaboration with cryptography experts. AICrypto includes three task types: multiple-choice questions (MCQs), capture-the-flag (CTF) challenges, and proof problems. These tasks span a wide spectrum of cryptographic skills, from conceptual knowledge to vulnerability exploitation and formal reasoning.

Specifically, MCQs test the model's factual memorization of fundamental cryptographic concepts. Proof problems go further by evaluating the model's ability to construct rigorous formal arguments, simulating academic-level reasoning. CTF challenges emphasize practical skills, requiring models to exploit vulnerabilities through source code analysis and numerical reasoning, mimicking real-world cryptographic attacks. Together, these components provide a multifaceted and in-depth evaluation of LLMs' cryptographic proficiency. 

An overview of AICrypto is shown in Figure~\ref{fig:overview}, and Table~\ref{tab:comparison} highlights the difference between AICrypto and other benchmarks that include cryptography-related evaluations of LLMs.

Moreover, automating the evaluation of proof problems is challenging~\cite{hendrycksmath2021}, and manual scoring by human experts is prohibitively time-consuming. To address this, we collaborate with cryptography experts to develop detailed scoring rubrics and reference solutions for each problem, and adopt automated proof graders similar to~\citet{luong-etal-2025-towards}. Our automated graders achieve high correlation with human expert assessments (\Sref{sec:auto_score}), and the detailed rubrics and reference solutions we provide could serve as valuable resources for future research.

The main contributions of this work are as follows:

\begin{itemize}

    \item We introduce \textbf{AICrypto}, a benchmark designed to evaluate the cryptography capabilities of LLMs. Covering three task types, AICrypto assesses skills ranging from knowledge memorization to vulnerability exploitation and formal reasoning. To ensure data integrity and avoid contamination, all tasks are carefully curated and verified by cryptography experts.

    \item We evaluate the performance of 17 state-of-the-art LLMs on AICrypto, as shown in Figure~\ref{fig:llm_performance}, and conduct an in-depth analysis of their cryptographic capabilities, offering insights into the potential future research of LLMs in cryptography. While their performance on MCQs and proof problems already matches or even exceeds that of human experts, there is still considerable room for improvement in the more application-oriented CTF challenges.

    \item  Our in-depth case studies reveal interesting insights. For example, while current LLMs demonstrate strong memorization of basic cryptographic concepts, they still struggle with mathematical comprehension. In particular, they lack the ability to perform dynamic reasoning and accurate numerical analysis, which limits their effectiveness on more complex cryptographic tasks.

\end{itemize}

\begin{table*}[t]
    \centering
    \small
    \caption{Comparison of AICrypto with existing benchmarks on cryptography evaluation. All values indicate counts.}
    \label{tab:comparison}
    \renewcommand{\arraystretch}{1.3} 
    \begin{tabular}{lcccc}
\toprule
\textbf{Benchmark} &
\makecell{\textbf{Crypto CTF}\\ \textbf{Challenges}} & \makecell{\textbf{Other Crypto}\\\textbf{Tasks}} & \makecell{\textbf{Total Crypto}\\\textbf{Tasks}} & \makecell{\textbf{LLMs}\\\textbf{Evaluated}} \\
\midrule
        % AICryptoBench Row (Highlighted)
        \textbf{AICrypto (Ours)} & \textbf{150} & \textbf{135 MCQ + 30 Proof} & \textbf{315} & \textbf{17} \\
        \midrule
        % Other Benchmarks
        CyBench~\cite{zhang2025cybench} & 16 & -- & 16 & 8 \\
        NYUCTF~\cite{shao2024nyuctfbench} & 62 & -- & 62 & 5 \\
        \citet{shao2024empirical}& 4 & -- & 4 & 6 \\
        InterCode-CTF~\cite{yang2023intercode} & 15 & -- & 15 & 2 \\
        \bottomrule
    \end{tabular}
\end{table*}

\section{Benchmark Creation}
We collaborate with professional cryptography experts to develop AICrypto, a comprehensive benchmark comprising three task types: multiple-choice questions (MCQs), capture-the-flag (CTF) challenges, and proof problems. This section provides a detailed overview of our expert contributors and tasks.

\subsection{Expert Panel}

To ensure the quality and reliability of our benchmark, we assemble and work extensively with a panel of domain experts with strong backgrounds in cryptography and cybersecurity. The team includes: (1) A tenure-track assistant professor specializing in cryptography, who teaches undergraduate- and graduate-level cryptography courses at one of the most prestigious computer science programs worldwide (we do not reveal its identity for anonymity). (2) Four Ph.D. students specializing in cryptography supervised by the assistant professor. (3) Two undergraduate students majoring in cybersecurity. We defer the detailed roles or contributions of each expert to Appendix~\ref{app:human_expert_details}.

\subsection{Multiple-Choice Questions (MCQs)}

The MCQ task is designed to assess the target model's understanding of fundamental cryptographic concepts. It consists of 135 questions, including 118 single-answer and 17 multiple-answer items. The questions are carefully curated from reputable educational sources, including cryptography exam papers from leading universities (e.g., Stanford, UCSD, UC Berkeley, MIT, and National Taiwan University), as well as public practice sets from online platforms such as \url{https://www.sanfoundry.com/} and \url{https://www.studocu.com/}. 

To ensure high assessment quality and prevent data contamination, we manually verify each question and rewrite all questions and options. For instance, in calculation-based questions, our experts modify numerical values, rephrase the text, and randomize answer choices. For flawed or ambiguous questions, we consult human experts to refine the content and ensure clarity and accuracy. For more details, please refer to Appendix~\ref{app:data_con}.

Figure~\ref{fig:overview} details the scope and distribution of questions across these domains. An example question is shown in Figure~\ref{fig:example_question}.

\subsection{Capture-the-Flag (CTF) Challenges}
Capture-the-flag (CTF) competitions are professional contests designed for cybersecurity practitioners. A typical cryptographic CTF challenge provides participants with the source code of a vulnerable encryption algorithm and its corresponding output. The objective is to identify and exploit the vulnerabilities to recover the original plaintext, typically the flag. 

Unlike proof problems and MCQs, CTF challenges closely mirror real-world attack scenarios and demand practical exploitation skills. While CTF-style tasks have been adopted to evaluate LLMs' cybersecurity capabilities~\citep{shao2024nyuctfbench, zhang2025cybench}, prior efforts are limited by a narrow selection of crypto-focused challenges and inconsistent quality standards.

\begin{figure*}[t]
  \centering
  \includegraphics[width=0.83\linewidth]{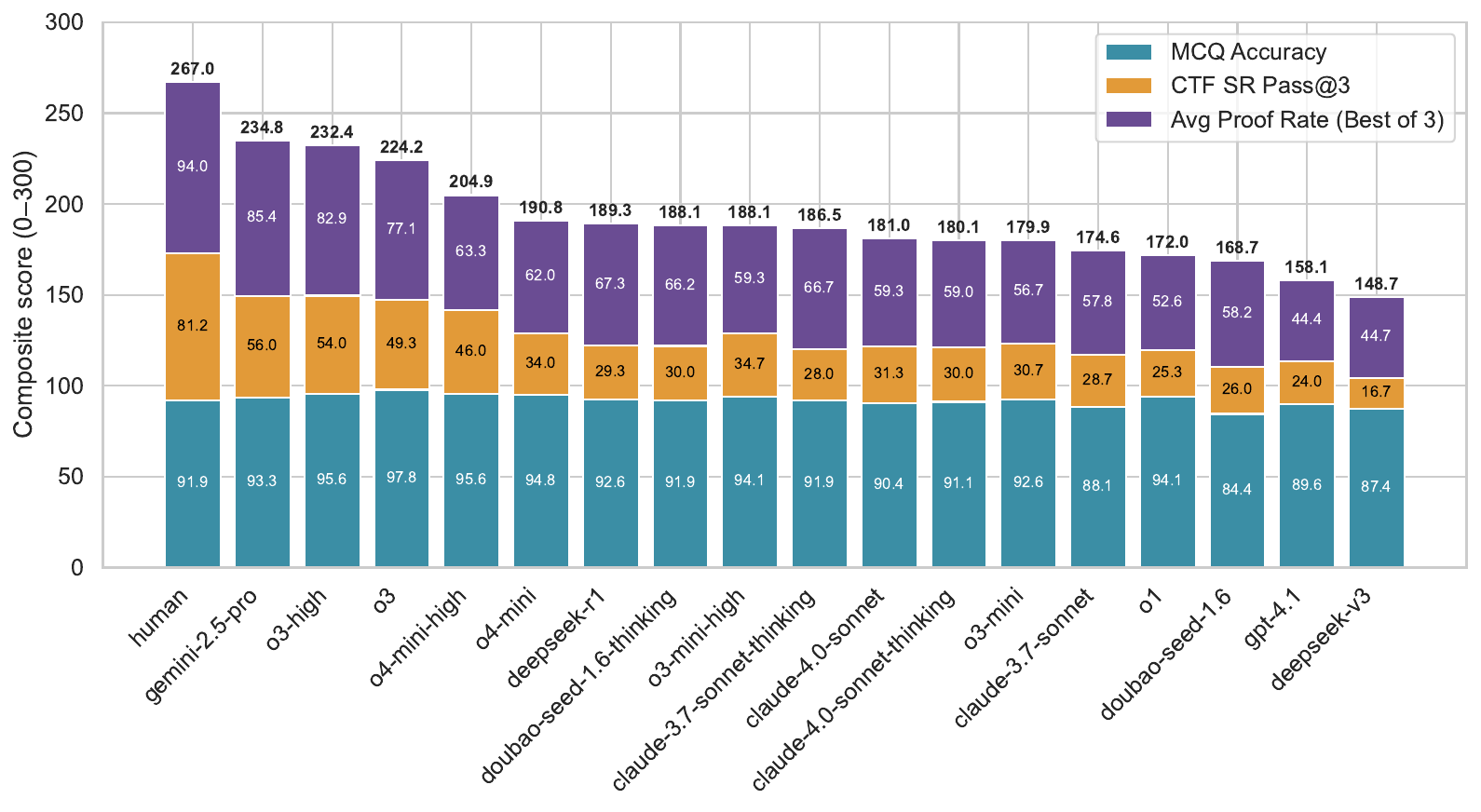}
  \vspace{-3mm}
  \caption{Comparison of LLMs' performance on AICrypto. For each model (ordered left-to-right by descending composite score), MCQ accuracy (teal), CTF success rate pass@3 (orange), and average proof scoring rate (purple) are stacked to yield the composite score.}
  %, with numeric labels denoting individual and overall scores.}
  \vspace{-1mm}
  \label{fig:llm_performance}
\end{figure*}

\begin{table*}[t]
\centering
\small
\caption{Human expert effort involved in benchmark creation (total person-hours, estimated).}
\label{tab:human_effort}

\begin{tabular}{@{} l l c c @{}}
\toprule
\textbf{Benchmark Part} & \textbf{Task} & \textbf{\# Experts} & \textbf{Time (Hours)} \\
\midrule
\multirow{2}{*}{MCQs} 
& Collection, rephrasing, and verification              & 3 & 120 \\
& Human baseline evaluation                & 3 & 36  \\
\midrule
CTF 
& Collection (challenges and solutions), verification 
                                          & 3 & 300 \\
\midrule
\multirow{2}{*}{Proof} 
& Proof problem creation                   & 1 expert + TAs & 50  \\
& Scoring criteria design, reference solutions, and grading LLM answers 
                                          & 4 & 40  \\
\midrule
\textbf{Total} & -- & -- & \textbf{546} \\
\bottomrule
\end{tabular}
\end{table*}

As shown in Figure~\ref{fig:overview}, AICrypto contains 150 CTF challenges across 5 categories. To ensure high quality, we collect challenges from well-established professional competitions, including \href{https://plaidctf.com/}{Plaid
 CTF} (organized by the CMU team), \href{https://2024.uiuc.tf/}{UIUCTF}
 (organized by the UIUC team), \href{https://ctf.dicega.ng/}{DiceCTF}
, and \href{https://cr.yp.toc.tf/}{CryptoCTF}
. To reduce the risk of data contamination, over 90\% of the challenges (137 out of 150) are sourced from 2023 or later. All challenges are carefully reviewed by human experts for both quality and correctness. For more details, please refer to Appendix~\ref{app:data_con}.

Figure~\ref{fig:ctf_example} illustrates a CTF challenge from AICrypto, originally featured in \href{https://ctftime.org/event/2126/}{BlueHens CTF 2023}. In a standard setup, human participants receive two files: \texttt{main.py} and \texttt{output.txt}. The file \texttt{main.py} implements an RSA-based encryption scheme that contains a known vulnerability (common modulus attack), while \texttt{output.txt} provides the corresponding ciphertext. The goal is to recover the original plaintext variable \texttt{msg}.

For the evaluated LLM, we also provide a \texttt{helper.py} or \texttt{helper.sage} script to assist with loading the data when needed. This is necessary because some challenge outputs are very long and may exceed the LLM's context window. The helper script allows LLMs to load truncated versions of these outputs. Further details are provided in Appendix~\ref{app:ctf_year}.
\vspace{-3mm}

\paragraph{Agent-based framework for CTF challenges.}
\label{sec:agent}

Solving CTF challenges typically requires writing programs that exploit vulnerabilities in cryptographic algorithm implementations or their underlying principles to recover the flags. As the tasks often involve complex large-number computations, an area where current LLMs struggle~\citep{yang-etal-2025-evaluating}. To address this, we adopt an agent-based evaluation framework~\citep{yao2023react, shao2024nyuctfbench, zhang2025cybench} in which the LLM functions as an autonomous agent that interacts with the environment.

Figure~\ref{fig:example_agent_output} shows how our agent works. We provide the task description, development environment specifications, and expected response format in the system and initial prompts. These prompts define the agent's goals and the set of permissible actions. During each interaction round, the model generates a response, from which we extract a single action, such as executing a command or creating a file. The environment then returns feedback, such as the output of the executed command or confirmation of a file being created. Through this iterative loop, the agent incrementally works toward recovering the flag. For more details, please refer to Appendix~\ref{app:ctf_framework_detail}.

\subsection{Proof Problems}
Proof problems are widely used in educational assessments, as they provide a deeper evaluation of a student's understanding than multiple-choice questions. Solving these problems requires a strong grasp of cryptographic concepts and solid logical reasoning skills. 

We select 30 cryptographic proof problems from assignments and exams used in cryptography courses at one of the most
prestigious computer science programs worldwide, from 2023 to 2025. For anonymity, we do not disclose the department's identity. These problems are thoughtfully handcrafted by the assistant professor in our expert panel and have never been publicly released online, which helps to prevent data contamination.

While these problems come from course materials, they are more challenging than typical undergraduate exercises. Many of them require reasoning similar to that needed in cryptographic research. Performance on these problems may indicate whether LLMs can assist with research-level reasoning in cryptography, which is why we include them in our benchmark.

As shown in Figure~\ref{fig:overview}, the 30 problems span core topics in cryptography: the foundation of cryptography (including one-way functions and hardcore functions, FUN), pseudorandomness (PR), and encryptions (ENC).

To ensure a rigorous evaluation and enable automated scoring, our human experts have manually designed scoring criteria and reference answers for every problem. This process requires substantial effort due to the unique challenges posed by each problem. Figure~\ref{fig:example_proof_problem} shows an example proof problem and its corresponding scoring criteria. We describe the automatic evaluation procedure in detail in ~\Sref{sec:auto_score}.  The complete set of problems is available in our anonymous repository for examination.

\subsection{Human Expert Effort}
The creation of AICrypto relies heavily on collaboration with human experts, who devote substantial time and effort to the project. Table~\ref{tab:human_effort} summarizes their contributions. In total, our human experts contribute an estimated \textbf{540+ hours} of work to build AICrypto.

The most time-consuming component is the CTF challenges, as we need to ensure that the core of each challenge tests cryptographic vulnerabilities rather than simple encryption tricks. This component is also the largest in scale, with 150 challenges. Additionally, we provide solutions for each challenge, which requires considerable time, as many challenges do not have publicly available official solutions.

The next component is the MCQs, where most of the time is spent on collection and rephrasing. The final component is the proof problems. All problems result from discussions and careful design by the assistant professor and teaching assistants. Our human experts provide reference solutions for every problem and review part of the LLM responses, covering 18 problems across 17 models (306 samples).

\begin{figure}[t]
\centering
\begin{mdframed}[linewidth=1pt, linecolor=black!30, backgroundcolor=lightgreenbin]
\small
\textbf{Example Multiple-Choice Question}

Given an RSA public key $(N,e)$ and the factorization $N = pq$, how can the secret key $d$ be computed?

\textbf{Options:}
\begin{enumerate}[label=\Alph*.]
    \item $d = e^{-1} \bmod \varphi(N)$, where $\varphi(N) = (p - 1)(q - 1)$ 
    \item $d = e^{-1} \bmod (N - 1)$
    
    \item $d = ep \bmod (q - 1)$ 
    % \item $d = p^{-1} \bmod q$
    % \item None of the above.
\end{enumerate}
\end{mdframed}
\caption{An example multiple-choice question from AICrypto. A is the correct option. Options D and E are omitted to save space.}
\vspace{-5mm}
\label{fig:example_question}
\end{figure}

\begin{figure*}[t]
  \centering
  \includegraphics[width=0.83\linewidth]{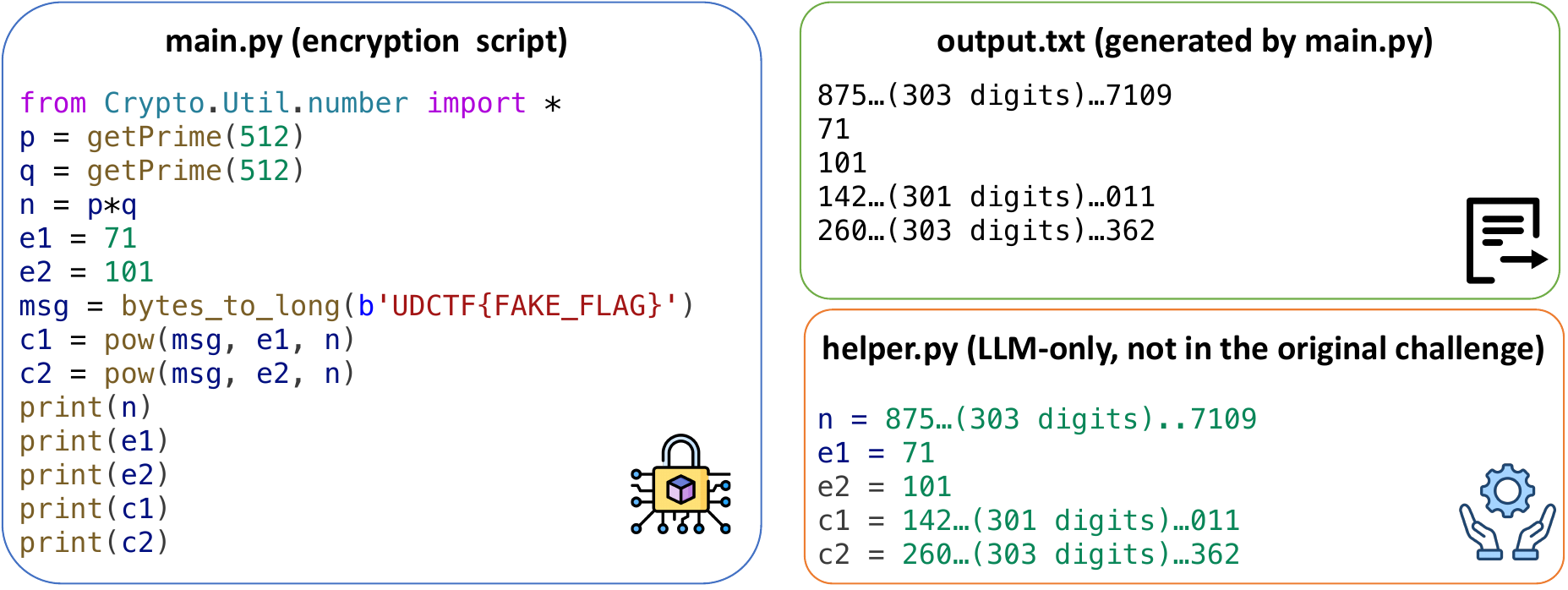}
  
  \caption{An example of a CTF challenge from AICrypto. Due to space constraints, only a portion of \texttt{output.txt} is shown. The marker ``(303 digits)'' indicates that 303 digits have been omitted.
}

  \label{fig:ctf_example}
\end{figure*}

\begin{figure*}[t]
\centering
\includegraphics[width=0.83\linewidth]{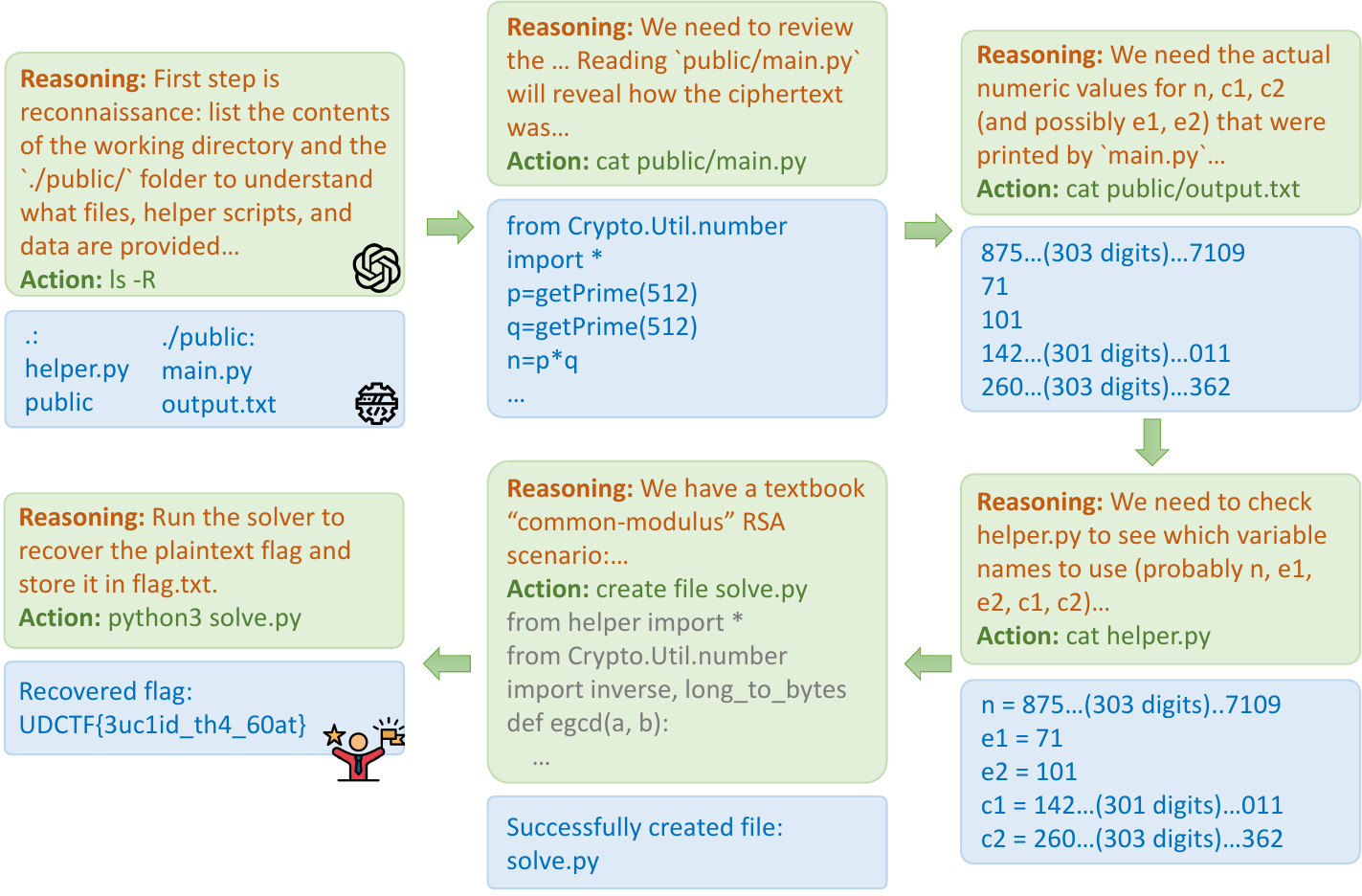}
\caption{A successful challenge-solving process by o3-high. The challenge corresponds to the one shown in Figure~\ref{fig:ctf_example}. For clarity, some model outputs and formatting details are omitted. The green box indicates the model's output, while the blue box represents feedback from the environment. The model correctly identifies the RSA vulnerability of common-modulus and successfully writes a script to recover the flag.}
\label{fig:example_agent_output}
\end{figure*}

\section{Experimental Setup}

In this section, we describe the experimental setup for each task. Details on how we obtain human expert performance are provided in Appendix~\ref{app: human_performance_details}.

\subsection{Models} 
We evaluate the performance of 17 models on AICrypto, including the following from the OpenAI series: {o3-high}, {o3}~\citep{o3o4mini}, {o4-mini-high}, {o4-mini}~\citep{o3o4mini}, o3-mini-high, o3-mini~\citep{o3-mini}, o1~\citep{o1}, and gpt-4.1~\citep{gpt-4.1}. From the Anthropic series, we include: claude-sonnet-4-thinking, claude-sonnet-4~\citep{claude4}, claude-sonnet-3.7-thinking, and claude-sonnet-3.7~\citep{claude3.7}. We also evaluate gemini-2.5-pro~\citep{gemini} from Google, the Deepseek models deepseek-v3~\citep{liu2024deepseek} and deepseek-r1~\citep{guo2025deepseek}, and the Doubao models doubao-seed-1.6~\citep{seed1.6} (unable thinking mode) and doubao-seed-1.6-thinking. All models are evaluated using their default settings. For detailed information on model versions and maximum tokens, please refer to Appendix~\ref{app:model}.

\subsection{Multiple-Choice Questions} 

For each multiple-choice question, we conduct a single-turn conversation to obtain the model's response. The model is instructed to follow a specific output format: it first provides an analysis of the question, then presents its final answer based on that analysis. We extract the answer by parsing the model's output. For the complete prompt, please refer to Appendix~\ref{app:prompt-mcq}.
\vspace{-3mm}
\paragraph{Metric.} We use the \textit{accuracy rate} as the evaluation metric for MCQs, calculated by dividing the number of correct answers by the total number of questions, with values ranging from 0 to 1. 
\vspace{-3mm}

\subsection{CTF Challenges} 
As detailed in~\Sref{sec:agent}, the LLM agent solves each CTF challenge through a multi-round interaction with the environment. We cap the conversation at 100 turns (i.e., 100 actions). The attempt is deemed a success only if the agent retrieves the correct flag within those 100 turns; if it exhausts the limit or opts to give up earlier, the attempt is recorded as a failure. For the complete prompts, please refer to Appendix~\ref{app:prompt-ctf}.
\vspace{-3mm}

\paragraph{Metric.} Following the pass@k metric commonly used in code generation tasks \citep{NEURIPS2019_7298332f}, we allow each LLM three independent attempts per challenge. If any attempt succeeds, we consider the task solved successfully; otherwise, it is marked as a failure. We use the \textit{success rate pass@3} as the metric for evaluating LLM performance.
\vspace{-3mm}

\subsection{Proof Problems}
\label{sec:auto_score}
Because proof problems come from both exams and homework, we adopt slightly different evaluation protocols. For exam problems, the model tackles each exam through a multi-round dialog: in each round, it answers exactly one problem and continues until it completes the entire exam. The full dialog history remains visible throughout, so the model can reuse earlier results when needed. In contrast, for homework problems, we evaluate each problem independently. For every problem, the model produces two sections, \textit{Analysis} and \textit{Proof}, and we grade only the \textit{Proof} section. The complete prompts are provided in Appendix~\ref{app:prompt-proof}.
\vspace{-3mm}
\paragraph{Automatic evaluation.} We use two of the most powerful models, gpt-5.1~\citep{gpt-5.1} and gemini-3-pro-preview~\citep{gemini-3-pro}, as grader models for the automated evaluation of proof problems. Each grader model receives the proof problem, scoring criteria, reference answer, and the answer being evaluated as input, and outputs its reasoning and a final score. For each answer, each grader model scores it independently three times, producing six scores in total, and we use their average as the final score. 

We also measure the correlation between human scores and LLM scores on 306 samples (18 problems $\times$ 17 models), the Pearson correlation between human experts and LLM scores is\textbf{ 0.9025 ($p < 10^{-10}$)}, and the Spearman correlation is \textbf{0.8973 ($p < 10^{-10}$)}. These high correlations demonstrate the effectiveness of our automatic grading strategy. We also evaluate other strategies, which we describe in detail in Appendix~\ref{app:auto_proof}, and we provide the grading prompts in Appendix~\ref{app:prompt-auto}.
\vspace{-3mm}

\paragraph{Metric.} 
Similarly, drawing inspiration from pass@k in code generation~\cite{NEURIPS2019_7298332f}, we evaluate each model three times and report the best total score across these three runs. Each run answers 30 problems, each worth 5 points, for a maximum of 150 points. We divide the best total score by 150 to obtain the final best-of-three score, which ranges from 0 to 1.

\begin{figure*}[!h]
\centering
\begin{mdframed}[linewidth=1pt, linecolor=black!30, backgroundcolor=lightbluebin]
\small
\textbf{Example Proof Problem and the Corresponding Scoring Criteria}

\paragraph{Exam 1, Problem 2 (5 points).}Show that there is no universal hardcore bit. 
In more detail, show that for every $n\in\mathbb{N}$, there is no deterministic function $h: \{0,1\}^n \to \{0,1\}$ such that for any polynomial $p$, \emph{any} one-way function $f: \{0,1\}^n \to \{0,1\}^{p(n)}$, $h$ is a hardcore bit for $f$. 

\paragraph{Scoring Criteria.}
\begin{description}[leftmargin=2.5em,labelsep=0.5em]

  \item[\textbf{2 points}]%
  For a universal hardcore bit $h(x)$, give a correct construction of a one-way
  function (such as $(g(x), h(x))$) such that $h(x)$ is \emph{not} a hardcore bit
  of this one-way function.

  \item[\textbf\^{}{2 points}]%
  Prove that the new function is actually one-way, including:
  \begin{description}[leftmargin=3em,labelsep=0.5em]
    \item[\textbf{1 point}]%
    Assume that it is not one-way, and correctly construct an algorithm that can
    be used to invert the original one-way function.

    \item[\textbf{1 point}]%
    Analyze the success probability of inverting the original one-way function.
  \end{description}

  \item[\textbf{\^{}1 point}]%
  Show that $h(x)$ is not a hardcore bit of the new one-way function.

\end{description}

\end{mdframed}
\vspace{-3mm}
\caption{Example proof problem and scoring rubric from AICrypto. \^{} indicates a parallel rule.}
\label{fig:example_proof_problem}
\end{figure*}

Finally, we calculate the composite score based on the results of the three tasks. The total composite score is 300 points, with each task contributing up to 100 points. A score of 1 point corresponds to a 1 percent accuracy, success rate, or scoring rate in the respective task.

\section{Result and Analysis}
\subsection{Result Overview}

Figure~\ref{fig:llm_performance} presents the overall performance of LLMs compared to human experts on AICrypto. The results reveal that these models fall into several performance tiers. \textbf{The three top-performing models are gemini-2.5-pro, o3-high, and o3, with scores around 230.} Among them, \textbf{gemini-2.5-pro achieves a high score of 234.8}, ranking first among all models and second only to human experts, who score 267.0. The second tier consists of models scoring between 180 and 220, representing the most (8 out of 17 models). The remaining 6 models score below 180 (60\%), and these are mainly earlier reasoning models and general models. Overall, reasoning models consistently outperform general models, and increased reasoning effort leads to better results.

A closer look at task-specific results reveals a clear pattern. \textbf{The most advanced LLMs surpass human experts on multiple-choice questions, and approach human performance on proof problems, but perform much worse on CTF challenges.} This suggests that while the top models have mastered fundamental concepts and reasoning skills in cryptography, they still struggle with real-world problem-solving. In the following sections, we provide a more detailed task-specific analysis. For additional results, please refer to Appendix~\ref{app:add_results}.

\subsection{Detailed Results on Different Tasks}
\begin{figure*}[t]
  \centering
  \includegraphics[width=0.83\linewidth]{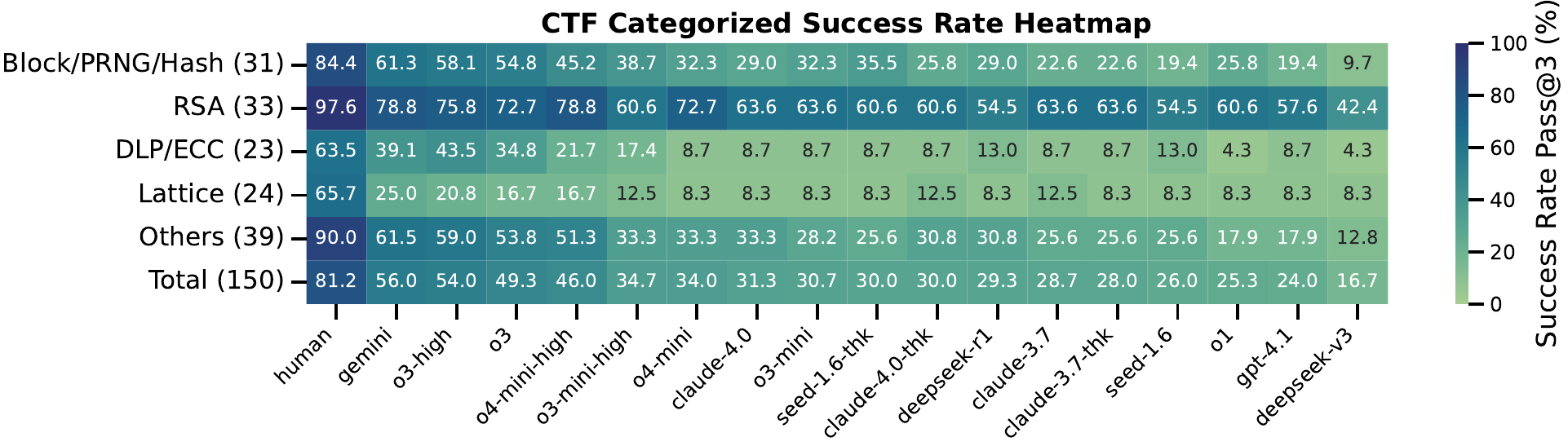}
  \caption{Heatmap of model and human expert success rate across different categories of \textbf{CTF challenges}. The y-axis labels indicate the challenge categories along with their corresponding counts. To save space, we abbreviate some model names. For example, gemini refers to gemini-2.5-pro.}
  \vspace{-3mm}
  \label{fig:ctf_heatmap}
\end{figure*}

\paragraph{Multi-choice questions.}

Figure~\ref{fig:mcp_heatmap} presents the accuracy of 17 LLMs and 3 human experts across 5 subcategories of MCQs. The o3 model makes only 3 errors out of 135 questions, reaching an overall accuracy of 97.8\% and achieving perfect scores in \emph{symmetric} and \emph{misc}. o4-mini-high and o3-high follow closely, clustering just below 96\%. Even the lowest-performing model, doubao seed-1.6, achieves a solid 84.4\%. The best human expert attains an accuracy of 94.1\% (127/135), which is strong but still below the state-of-the-art models. This demonstrates that current LLMs possess a solid understanding of basic cryptographic concepts and knowledge.
\vspace{-3mm}

\paragraph{CTF challenges.}
Figure~\ref{fig:ctf_heatmap} shows a category-level successful rate heatmap for 17 LLMs and a panel of human experts on CTF challenges (human performance calculated from a subset of 100 challenges). Human experts lead with an average success rate of 81.2\%, while the best-performing models, gemini-2.5-pro and o3-high, reach only 56.0\% and 54.0\%, respectively. The second tier, including o3 and o4-mini-high, achieves 49.3\% and 46.0\% respectively. Performance drops steeply among the remaining models, all of which have a success rate below 35\%. These results highlight a persistent 25–30 percentage point gap between top LLMs and human experts. 

Across all model families, larger models or those configured with greater reasoning effort consistently outperform their smaller or less intensive counterparts. For example, o3-high outperforms o3, which in turn surpasses o3-mini. This performance hierarchy mirrors trends observed in other domains~\citep{balunovic2025matharena, qiu2025phybenchholisticevaluationphysical}.

Overall, LLMs perform well on challenges based on well-known cryptographic vulnerabilities, such as the common-modulus attack in RSA, but they continue to struggle with tasks like lattice-based problems that demand advanced mathematical reasoning and creativity.
\vspace{-3mm}

\begin{figure*}[t]
  \centering
  \includegraphics[width=0.83\linewidth]{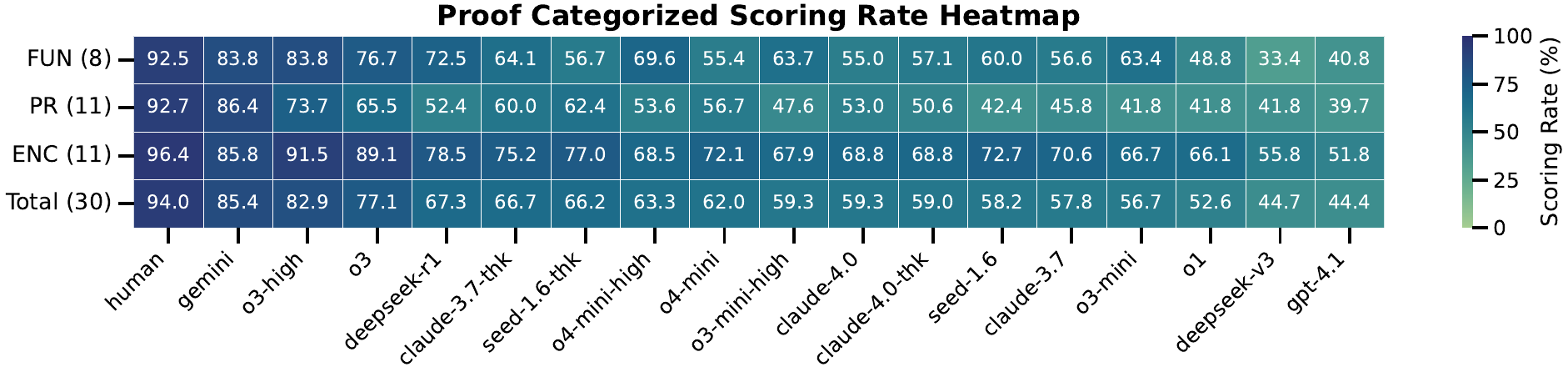}
  \caption{Heatmap of model and human expert scoring rates across different categories of \textbf{proof problems}. The y-axis labels indicate the problem categories along with their corresponding counts. To save space, we abbreviate some model names. For example, gemini refers to gemini-2.5-pro.}
  \vspace{-3mm}

  \label{fig:proof_heatmap}
\end{figure*}

\paragraph{Proof problems.} 
Figure~\ref{fig:proof_heatmap} shows category-level scoring rates for 17 LLMs compared with human experts on proof problems. Human experts hold a slight advantage with an average score of 94.0\%. The best models also perform strongly, with gemini-2.5-pro at 85.4\% and o3-high at 82.9\%, both close to human level. Half of the models, 9 out of 17, score below 60\%. The proofs provided by the models are available in Appendix~\ref{app:proof}.

\subsection{Failure Case Analysis}
\label{sec:error-study}
In this section, we analyze and discuss the reasons why LLMs fail to solve certain tasks in AICrypto. All conclusions are based on manual inspection conducted by human experts. Detailed cases are provided in Appendix~\ref{app:detailed_fa_case}.
\vspace{-4mm}

\begin{figure*}[t]
  \centering
  \includegraphics[width=0.83\linewidth]{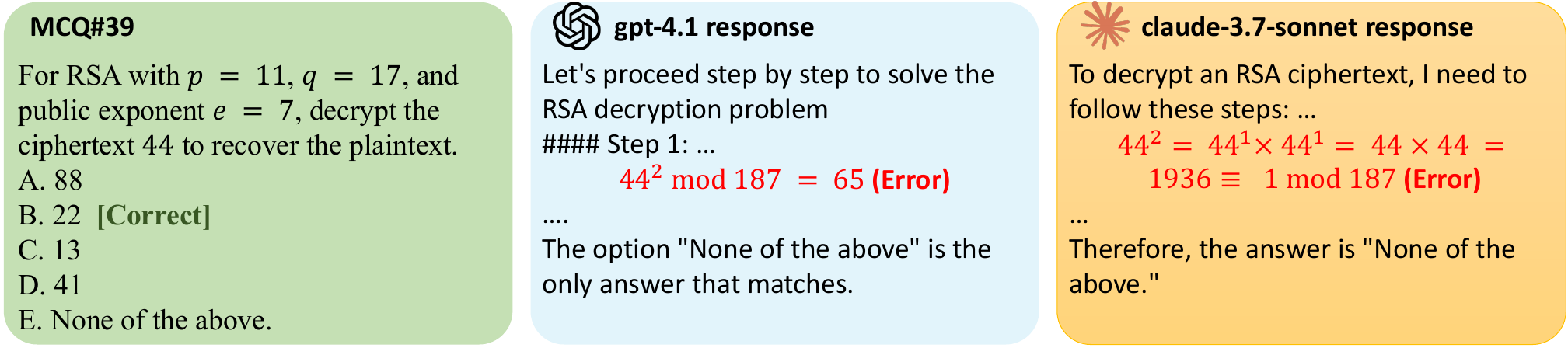}
  \caption{An example of a calculation error made by LLMs.}
  \vspace{-4mm}
  \label{fig:case1}
\end{figure*}

\paragraph{Inaccuracy of mathematical computation.}

LLMs exhibit certain deficiencies in performing precise numerical calculations~\citep{yang-etal-2025-evaluating}. These computational errors persist across various models, indicating a systematic difficulty in handling even relatively simple arithmetic operations.  As shown in Figure~\ref{fig:case1}, gpt-4.1 and claude-3.7-sonnet incorrectly compute the basic modular exponentiation $\left(44^{2}\right) \bmod 187$.

 \paragraph{Excessive reliance on pattern matching over analysis.} 
In CTF challenges, LLMs tend to perform well on straightforward mathematical tasks, such as solving equations or inverse problems, as well as brute-force searches. However, they struggle with tasks that require dynamic or recursive reasoning, particularly those involving logical or numerical analysis. As a result, models often default to mechanically applying familiar attack patterns, rather than engaging in the deeper analytical thinking necessary for tackling novel or complex cryptographic problems.
\vspace{-3mm}

\paragraph{Limitations in mathematical comprehension.} Current LLMs exhibit significant limitations in their ability to understand and reason about complex mathematical concepts. Their proof writing suggests that LLMs may primarily mimic the syntactic structure of proof languages provided by humans, without truly underlying mathematical principles, such as the precise meaning of ``one-way function", ``pseudorandom", ``computationally indistinguishable'', or something else. 
\vspace{-3mm}

\paragraph{Deficiencies in rigorous mathematical proof-writing.} LLMs often struggle to produce mathematically rigorous and complete proofs. Their constructions often contain logical gaps or omit essential technical details. In some cases, they generate proofs that appear correct at first glance but reveal critical flaws upon closer examination. Such issues make it particularly difficult for human graders to detect errors, reducing the reliability of using these tasks to assess LLMs' proof-generation capabilities.

\section{Related Work}
\paragraph{Benchmarking cybersecurity capabilities of LLMs.}Cybersecurity is a critical research area, and as LLMs' capabilities advance, several efforts have emerged to evaluate their proficiency in this domain. Earlier work primarily focuses on CTF-style tasks, exemplified by benchmarks like Cybench~\citep{zhang2025cybench} and NYU CTF Bench~\citep{shao2024nyuctfbench}. More recently, evaluations have shifted towards practical applications with benchmarks such as CVE-Bench~\citep{cvebench}, PentestGPT~\citep{PentestGPT}, BountyBench~\citep{zhang2025bountybenchdollarimpactai}, and CyberGym~\citep{wang2025cybergym}. However, these benchmarks include cryptography only as a minor component, and the quality of cryptographic problems is often limited. For example, Cybench contains 40 CTF challenges in total, while NYU CTF Bench includes 52 crypto CTF challenges, some of which focus on miscellaneous encryption techniques rather than core cryptographic algorithms. Similarly, \citet{li2025cipherbank} uses decryption tasks to study LLM reasoning, but their tasks are restricted to classical cryptography.

\paragraph{Cryptography in AI.} Cryptography and its underlying principles have long played a vital role in artificial intelligence. For example, differential privacy~\citep{abadi2016deep}, homomorphic encryption~\citep{aono2017privacy}, and secure multi-party computation~\citep{knott2021crypten} are widely applied to protect privacy in machine learning and deep learning. Additionally, deep learning itself has been explored as a method to build desired cryptographic functionalities~\citep{gerault2025securely}. Beyond protective applications, cryptanalysis is employed to extract neural network models~\citep{carlini2020crypto, carlini2025polynomial}, while machine learning has emerged as a powerful tool for cryptanalysis~\citep{yu2018asiacrypt, li2023acmccs}. The recent rise of LLMs has further sparked interest in cryptographic applications in AI. For instance, some researchers explore the use of LLMs in cryptanalysis~\citep{maskey2025benchmarking}, while others draw on cryptographic inspiration to jailbreak LLMs~\citep{halawi2024covert, wang2024jailbreak}. As LLMs continue to improve, especially in mathematical reasoning and programming abilities, we anticipate a wave of innovative and surprising applications in the intersection of cryptography and AI. To save space, some related work is provided in Appendix~\ref{app:related_work}.

\section{Limitations}

\paragraph{Saturated performance on the multiple-choice questions.} AICrypto comprises multiple task types designed to evaluate different skills. The MCQs primarily assess basic concepts and knowledge, areas in which LLMs perform strongly. As a result, the LLM performance on the MCQs of AICrypto is nearing saturation. Nevertheless, we view this as only an initial step, as no prior work has systematically studied LLM capabilities in cryptography. We hope that our findings can inform future efforts toward developing more challenging cryptographic benchmarks. At the same time, there remains room for improvement on AICrypto's proof problems and CTF challenges, where current LLMs continue to struggle. 
\vspace{-4mm}
\paragraph{Limited exploration of agent frameworks.} Our evaluation focuses on the intrinsic capabilities of LLMs rather than agent system performance. We use a relatively simple agent framework only for CTF challenges (see~\Sref{sec:agent}) and rely on pure LLM outputs for the other tasks. We do not explore advanced agent frameworks such as CodeX~\citep{codex} or multi-agent collaboration~\citep{hong2023metagpt}, which could improve performance. Investigating these systems is left for future work.

\section{Conclusion}

We introduce AICrypto, a comprehensive benchmark for evaluating LLMs' cryptography capabilities, developed through extensive collaboration with strong experts. AICrypto includes 135 multiple-choice questions, 150 CTF challenges, and 30 proof problems. To improve the rigor and correctness, manual curation and expert verification are applied to all benchmark components. From the evaluation of 17 state-of-the-art LLMs, we find that leading models excel at knowledge memorization, basic vulnerability exploitation, and formal proof generation. The performance of top models matches or surpasses human experts in these areas. However, they continue to struggle with precise numerical analysis, deep mathematical reasoning, and multi-step planning required for complex tasks. These findings highlight both substantial potential and intriguing limitations of LLMs in cryptography, and we hope our work could provide insights for future research.

\section*{Acknowledgment}
We sincerely thank Wei Xu, Binyi Chen, Yixin Tao, Yu Yu, Yichen Wang, and Kaifeng Lyu for their valuable discussions and insightful comments that helped improve this work. This work was supported by the National Natural Science Foundation of China (No. 62176252) and the Shanghai Qi Zhi Institute Innovation Program.

\section*{Impact Statement}
This work evaluates the capabilities of LLMs in cryptography tasks. All experiments strictly follow ethical research principles. The tasks in this study are either publicly available or manually created by human experts. No sensitive or private data is used. Human expert evaluations are voluntary, conducted with informed consent, and participants face no risk beyond typical academic activities. 

The results aim to improve understanding of LLM capabilities in cryptography and cybersecurity, highlighting both strengths and limitations. We do not intend this study to promote the development or use of LLMs for malicious purposes, and all experiments take place in safe, controlled academic settings. The LLM usage statement is provided in Appendix~\ref{app:llm_usage}.


\bibliography{example_paper}
\bibliographystyle{icml2026}

%%%%%%%%%%%%%%%%%%%%%%%%%%%%%%%%%%%%%%%%%%%%%%%%%%%%%%%%%%%%%%%%%%%%%%%%%%%%%%%
%%%%%%%%%%%%%%%%%%%%%%%%%%%%%%%%%%%%%%%%%%%%%%%%%%%%%%%%%%%%%%%%%%%%%%%%%%%%%%%
% APPENDIX
%%%%%%%%%%%%%%%%%%%%%%%%%%%%%%%%%%%%%%%%%%%%%%%%%%%%%%%%%%%%%%%%%%%%%%%%%%%%%%%
%%%%%%%%%%%%%%%%%%%%%%%%%%%%%%%%%%%%%%%%%%%%%%%%%%%%%%%%%%%%%%%%%%%%%%%%%%%%%%%
\newpage
\appendix
\onecolumn
\definecolor{promptbg}{rgb}{0.97,0.97,0.97}
\definecolor{promptborder}{rgb}{0.7,0.7,0.7}
\captionsetup{hypcap=false}

\tcbset{
  promptstyle/.style={
    enhanced,
    colback=promptbg,
    colframe=promptborder,
    fonttitle=\bfseries,
    coltitle=black,
    sharp corners,
    boxrule=0.4pt,
    left=6pt,
    right=6pt,
    top=4pt,
    bottom=4pt,
    boxsep=2pt
  }
}

\lstset{
basicstyle=\rmfamily\normalsize,    
  breaklines=true,               
  breakatwhitespace=true,        
  columns=fullflexible,
  breakindent=0pt,
   literate=
    {'}{\textquotesingle}1
    {`}{\textasciigrave}1,   
}

\newpage

\section{LLM Usage Statement}
\label{app:llm_usage}
We use LLMs to help refine the language of this manuscript, including suggesting alternative wording and phrasing, checking grammar, and enhancing overall fluency and readability. All scientific content, ideas, analyses, and conclusions remain our own; the LLMs serve solely as tools to improve the presentation of the text.

\section{Related Work}
\label{app:related_work}
\paragraph{Benchmarking programming capabilities of LLMs.} Another closely related field is the evaluation of LLMs in programming. A substantial body of work investigates how to evaluate programming-related capabilities of LLMs. HumanEval~\citep{chen2021evaluating} is an early benchmark that systematically evaluates code generation performance using 164 hand-written programming problems. Building on this, LiveCodeBench~\citep{jain2025livecodebench} offers a comprehensive, contamination-free evaluation by continuously aggregating problems from various programming contests. Other efforts focus on evaluating coding abilities in real-world development scenarios, such as SWE-Bench~\citep{jimenez2023swe}, BigCodeBench~\citep{zhuo2025bigcodebench}, and NoFunEval~\citep{singhal2024nofuneval}. Additionally, TCGBench~\citep{cao2025can} explores LLMs’ capabilities in generating robust test case generators, providing a dual evaluation of their capabilities in both programming problem understanding and code understanding.

\section{CTF Challenge Details}
\label{app:ctf_year}
\subsection{Challenge Type and File Structure}

Cryptographic CTF challenges in AICrypto fall into two types: \textit{static} and \textit{dynamic}. Static challenges provide participants with all necessary files to recover the flag locally, while dynamic challenges involve interacting with a running server (e.g., \texttt{localhost:1337}) based on incomplete server-side code that withholds key information such as the flag. There are 73 static challenges and 77 dynamic challenges.

\begin{wrapfigure}{r}{0.5\textwidth}
    \centering
    \includegraphics[width=0.48\textwidth]{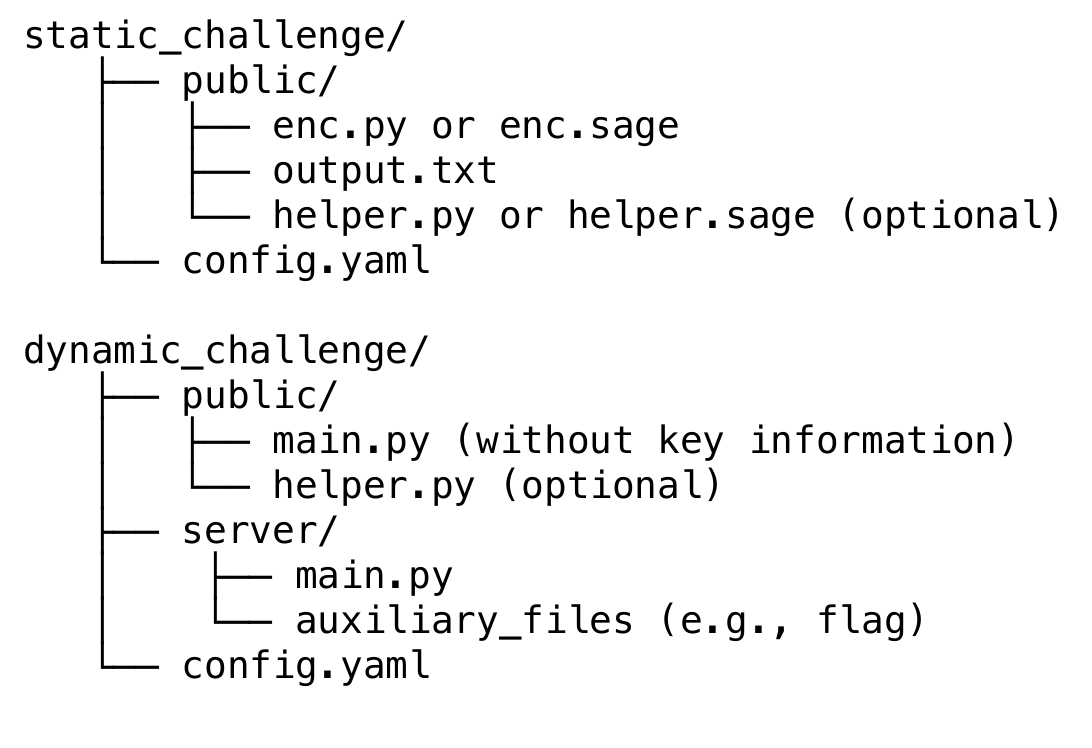}
    \caption{CTF challenge file structure.}
    \vspace{-5mm}
    \label{fig:chal_format}
\end{wrapfigure}

To support large-scale evaluation, we standardize each challenge into a unified file structure. Figure~\ref{fig:chal_format} illustrates the typical file layout for two types of challenges. Every challenge includes a \texttt{public} folder and a \texttt{config.yaml} file, while dynamic challenges add a \texttt{server} folder containing the server's launch script (\texttt{main.py}). Some files within the \texttt{public} folder may vary slightly across different challenges.
\vspace{-5mm}

\paragraph{Static challenges.} In static challenges, the \texttt{public} folder typically contains the encryption algorithm's source code and its corresponding output. These scripts are written in Python or SageMath, with the latter being an open-source mathematical software system built on Python. 

\paragraph{Dynamic challenges.} For dynamic challenges, the \texttt{public} folder contains a partial version of the server code, deliberately omitting key elements such as the flag. Before the LLM start the challenge, we will launch the \texttt{main.py} script from the \texttt{server} directory, expose it on a designated port, and provide the connection details to the LLM. To retrieve the actual flag, the LLM must analyze the available code and craft interactive scripts capable of communicating effectively with the running server.

\paragraph{Helper scripts.} Because some code and output files involve very large numbers or complex data, we provide a \texttt{helper.py} or \texttt{helper.sage} script\footnote{We provide helper scripts for 68 challenges, one per challenge. Among them, 48 are implemented as \texttt{helper.py} scripts and 20 as \texttt{helper.sage} scripts. Of these, 3 scripts are designed for dynamic challenges, while the remaining 65 support static challenges.} to assist LLMs in loading and processing the data. For instance, models can simply use \texttt{from helper import *} to access relevant variables. In addition, since some output files are very long and may exceed the model's context window, we truncate any output beyond 4096 characters and indicate the omission as shown in Figure~\ref{fig:ctf_example}. The presence of helper scripts ensures that this abbreviation does not hinder models from solving the problem.

\paragraph{Configuration.} We provide a configuration file named \texttt{config.yaml} for each challenge. As shown in Figure~\ref{fig:ctf-config}, this file records essential information including the category, correct flag, source, name, solution execution time and type of the challenge. Unlike prior CTF benchmarks, we omit original challenge descriptions and instead reformat all tasks into a unified structure, using standardized prompts to guide model behavior. 
For all challenges, models have access only to the \texttt{public} folder during evaluation, all other components remain hidden.

\subsection{CTF Challenge Year Distribution} Table~\ref{tab:ctf_chals_by_year} summarizes the number of CTF challenges by year. Most challenges originate from 2023 and beyond, with 102 problems from 2024 or later, accounting for 65\% of the dataset. This distribution highlights that our benchmark is primarily composed of recent and up-to-date challenges, ensuring relevance to the current cryptographic landscape.

\subsection{Solution} We provide solutions for each challenge in its corresponding folder to ensure that every challenge has a valid solution. These solutions are also made publicly available with the hope of offering useful insights and support for future research.

\begin{figure}[H]
\centering
\begin{tcolorbox}[promptstyle, title=An Example of CTF Configuration File]
category: RSA \\
flag: UDCTF\{3uc1id\_th4\_60at\} \\
from: blue-hens-2023 \\
name: RSA\_School\_3rd\_Grade \\
type: static\\
time: 300
\end{tcolorbox}
\caption{An example configure file for a CTF challenge.}
\label{fig:ctf-config}
\end{figure}

\begin{table}[h]
\centering
\caption{Distribution of CTF challenges by year.}
\label{tab:ctf_chals_by_year}
\begin{tabular}{|c|c|}
\hline
\textbf{Year} & \textbf{Number of CTF Problems} \\
\hline
2019 & 2 \\
\hline
2020 & 5 \\
\hline
2021 & 2 \\
\hline
2022 & 4 \\
\hline
2023 & 35 \\
\hline
2024 & 80 \\
\hline
2025 & 22 \\
\hline
\textbf{Total} & \textbf{150} \\
\hline
\end{tabular}

\end{table}

\definecolor{promptbg}{rgb}{0.97,0.97,0.97}
\definecolor{promptborder}{rgb}{0.7,0.7,0.7}
\definecolor{rea}{RGB}{0, 158, 115}    
\definecolor{ac}{RGB}{213, 94, 0}  
\definecolor{act}{RGB}{0, 114, 178}
\definecolor{res}{RGB}{204,121,167}
\definecolor{success}{RGB}{0,128,0}

\captionsetup{hypcap=false}

\tcbset{
  promptstyle/.style={
    enhanced,
    colback=promptbg,
    colframe=promptborder,
    fonttitle=\bfseries,
    coltitle=black,
    sharp corners,
    boxrule=0.4pt,
    left=6pt,
    right=6pt,
    top=4pt,
    bottom=4pt,
    boxsep=2pt
  }
}

% \lstset{
% basicstyle=\rmfamily\normalsize,    
%   breaklines=true,               
%   breakatwhitespace=true,        
%   columns=fullflexible,
%   breakindent=0pt,
%    literate=
%     {'}{\textquotesingle}1
%     {`}{\textasciigrave}1,   
% }

\lstdefinestyle{logcode}{
    language=Python,
    basicstyle=\ttfamily\small, 
    numberstyle=\tiny\color{gray}, 
    keywordstyle=\color{blue}, 
    commentstyle=\color{green!50!black}, 
    stringstyle=\color{orange},
    breaklines=true,
    frame=single,
    showstringspaces=false,
    tabsize=4
}

\lstdefinestyle{reasoning}{
    basicstyle=\rmfamily\normalsize\color{rea},
    numbers=none,
    identifierstyle=\color{rea},
    stringstyle=\color{rea},
    commentstyle=\color{rea},
    keywordstyle=\color{rea},
    emphstyle=\color{rea},
    alsoletter={\(\))},
    literate={
        {`}{{\color{rea}`}}1
        {'}{{\color{rea}'}}1
        {/}{{\color{rea}/}}1
        {.}{{\color{rea}.}}1
        {0}{{\color{rea}0}}1
        {1}{{\color{rea}1}}1
        {2}{{\color{rea}2}}1
        {3}{{\color{rea}3}}1
        {4}{{\color{rea}4}}1
        {5}{{\color{rea}5}}1
        {6}{{\color{rea}6}}1
        {7}{{\color{rea}7}}1
        {8}{{\color{rea}8}}1
        {9}{{\color{rea}9}}1
        % {*}{{\color{rea}*}}1
        % {(}{{\color{rea}{(}}}1
        % {)}{{\color{rea})}}1
        % {\^}{{\color{rea}{\^}}}1
        % {+}{{\color{rea}+}}1
        {-}{{\color{rea}-}}1
        % {=}{{\color{rea}=}}1
        {,}{{\color{rea},}}1
    },
    aboveskip=0pt,
    belowskip=0pt
}

\lstdefinestyle{action}{
    basicstyle=\rmfamily\normalsize\color{ac},
    numbers=none,
    identifierstyle=\color{ac},
    stringstyle=\color{ac},
    commentstyle=\color{ac},
    literate={
        {`}{{\color{ac}`}}1
        {'}{{\color{ac}'}}1
        {/}{{\color{ac}/}}1
        {.}{{\color{ac}.}}1
    },
    aboveskip=0pt,
    belowskip=0pt
}

\lstdefinestyle{action_content}{
    basicstyle=\rmfamily\normalsize\color{act},
    numbers=none,
    identifierstyle=\color{act},
    stringstyle=\color{act},
    commentstyle=\color{act},
    literate={
        {`}{{\color{act}`}}1
        {'}{{\color{act}'}}1
        {/}{{\color{act}/}}1
        {.}{{\color{act}.}}1
    },
    aboveskip=0pt,
    belowskip=0pt
}

\lstdefinestyle{response}{
  basicstyle=\rmfamily\normalsize\color{res},
    numbers=none,
    identifierstyle=\color{res},
    stringstyle=\color{res},
    commentstyle=\color{res},
    literate={
        {`}{{\color{res}`}}1
        {'}{{\color{res}'}}1
        {/}{{\color{res}/}}1
        {.}{{\color{res}.}}1
    },
    aboveskip=0pt,
    belowskip=0pt
}

\section{CTF Agentic Framework}
\label{app:ctf_framework_detail}
\paragraph{LLM agent output.} We format the agent output to support seamless interaction with the environment. The response is divided into three tagged sections: \texttt{\#\#\# Reasoning}, \texttt{\#\#\# Action}, and \texttt{\#\#\# Action Content}. The \texttt{Reasoning} section details the model’s thought process; the \texttt{Action} specifies one of four options: \texttt{command}, \texttt{create file <path>}, \texttt{verification}, or \texttt{give up}; and the \texttt{Action Content} provides the corresponding command or file contents. We require the model to save the recovered flag in \texttt{./flag.txt}. When the model requests verification, we compare the contents of \texttt{flag.txt} with the correct flag. If the model concludes the task is unsolvable, it may choose to give up, which marks the attempt as failed. Figure~\ref{fig:example_agent_res} shows an example of the agent's output.

\paragraph{Execution environment.} Drawing inspiration from typical CTF player setups, we configure the agent to operate within a controlled Ubuntu-based environment. Before each task begins, we copy the relevant \texttt{public} folder to a designated writable path. This is the only folder accessible at runtime, and the agent must create and execute all exploit scripts within it. To ensure controlled and secure execution, we prohibit directory changes (e.g., via \texttt{cd}) and enforce a blacklist of potentially unsafe commands, which return a ``permission denied'' message if attempted. In addition to Python and SageMath, the environment includes auxiliary tools such as \texttt{yafu}. More experimental details are provided in the Appendix~\ref{app:ctf_exp_env}.

\begin{figure}[t]
\centering
\begin{mdframed}[linewidth=1pt, linecolor=black!30, backgroundcolor=lightgray!20]
\textbf{An Example of Agent Output}

\#\#\# Reasoning

Run the solver script to recover the plaintext and automatically write the flag to \texttt{flag.txt}.
\\

\#\#\# Action

command
\\

\#\#\# Action Content

python solve.py

\end{mdframed}
\caption{An example of an agent's response template.}
\vspace{-3mm}
\label{fig:example_agent_res}
\end{figure}

\section{Data Contamination}
\label{app:data_con}
\paragraph{Multiple-choice questions.} To further assess contamination in the MCQs, we conduct a retrieval-based probe across all 135 questions. For each question, we retrieve the top-10 public candidates via DuckDuckGo\footnote{\url{https://duckduckgo.com/}} search and measure normalized n-gram overlap between each MCQ and its nearest retrieved source. 

\begin{table}[t]
    \centering
    \small
    \caption{Retrieval-based contamination analysis for the MCQs.}
    \label{tab:mcq_contamination}
    \renewcommand{\arraystretch}{1.2}

    \begin{tabular}{lc}
        \toprule
        \textbf{Metric} & \textbf{Value} \\
        \midrule
        Median 5-gram contamination & 0.0000 \\
        Mean 5-gram contamination & 0.0722 \\
        Median 7-gram contamination & 0.0000 \\
        Mean 7-gram contamination & 0.0368 \\
        5-gram contamination $> 0.3$ & 12 / 135 \\
        5-gram contamination $> 0.5$ & 4 / 135 \\
        7-gram contamination $> 0.3$ & 6 / 135 \\
        7-gram contamination $> 0.5$ & 3 / 135 \\
        \bottomrule
    \end{tabular}
\end{table}

The results are shown in Table~\ref{tab:mcq_contamination}. Both median containment scores are 0.0, indicating that verbatim overlap with public sources is minimal for most items. The small subset with higher overlap (at most 12 out of 135 at the 0.3 threshold) is consistent with our rewriting process, which targets surface-form leakage.

\paragraph{Capture-the-flag challenges.} For CTF challenges, the main contamination risk comes from the release of official solutions. We analyze all 85 challenges with official solutions (w/ OS) available online (56.67\% of the full set) and compare models' success rates (SR) against the remaining 65 challenges without official solutions (w/o OS).

\begin{table*}[t]
    \centering
    \footnotesize
    \setlength{\tabcolsep}{5pt}
    \caption{Comparison of model success rates on CTF challenges with official solutions available online (w/ OS) and without official solutions (w/o OS).}
    \label{tab:official_solution_contamination}
    \renewcommand{\arraystretch}{1.15}

    \begin{tabular}{lcc}
        \toprule
        \textbf{Model} & \textbf{Success Rate w/ OS} & \textbf{Success Rate w/o OS} \\
        \midrule

        gemini-2.5-pro & 47.06\% (40/85) & 67.69\% (44/65) \\
        o3-high & 47.06\% (40/85) & 63.08\% (41/65) \\
        o3 & 42.35\% (36/85) & 58.46\% (38/65) \\
        o4-mini-high & 38.82\% (33/85) & 55.38\% (36/65) \\
        o3-mini & 27.06\% (23/85) & 35.38\% (23/65) \\
        claude-4.0-sonnet & 25.88\% (22/85) & 38.46\% (25/65) \\
        o3-mini-high & 25.88\% (22/85) & 46.15\% (30/65) \\
        o4-mini & 25.88\% (22/85) & 44.62\% (29/65) \\
        deepseek-r1 & 23.53\% (20/85) & 36.92\% (24/65) \\
        claude-4.0-sonnet-thinking & 22.35\% (19/85) & 40.00\% (26/65) \\
        doubao-seed-1.6-thinking & 21.18\% (18/85) & 41.54\% (27/65) \\
        claude-3.7-sonnet & 21.18\% (18/85) & 38.46\% (25/65) \\
        doubao-seed-1.6 & 20.00\% (17/85) & 33.85\% (22/65) \\
        o1 & 20.00\% (17/85) & 32.31\% (21/65) \\
        gpt-4.1 & 20.00\% (17/85) & 29.23\% (19/65) \\
        claude-3.7-sonnet-thinking & 20.00\% (17/85) & 38.46\% (25/65) \\
        deepseek-v3 & 10.59\% (9/85) & 10.59\% (9/65) \\

        \bottomrule
    \end{tabular}
\end{table*}

The results are shown in the Table~\ref{tab:official_solution_contamination}. Across all models, the presence of official solutions does not correspond to higher success rates. Models tend to perform worse on problems with official write-ups. This suggests that data contamination has not meaningfully inflated model performance.

\section{Experiment Setup}

\subsection{Human Expert Performance Evaluation}
\label{app: human_performance_details}
We include the performance of strong human experts as a comparison during evaluation. The following describes how we obtain their performance on different tasks:

\paragraph{Multiple-choice questions.} To establish a human expert performance baseline, we recruit three doctoral students specializing in cryptography from a top university. They complete the multiple-choice section as an open-book exam, using only a designated reference textbook\footnote{\url{https://www.uoitc.edu.iq/images/documents/informatics-institute/Competitive_exam/Cryptography_and_Network_Security.pdf}}. The allotted time for answering is limited to 12 hours with breaks. Participants may consult only the reference book and use a non-programmable calculator. We do not permit the use of calculators for LLMs, as the calculations required for the MCQs are minimal.  We report the average accuracy achieved by the three experts.

\paragraph{Capture-the-flag challenges.} We estimate human expert performance using recorded scoreboards from CTF competitions. Specifically, we treat the top 10 participants in each competition as human experts and use their success rates to establish the human baseline. Since not all competitions provide detailed rankings, we collect the available data for 100 challenges and use this subset to as a proxy for average expert-level human performance.

\paragraph{Proof problems.} 
Because our proof problems come directly from real assignments and exams, we select the top five scores from each source as the expert scores.

\subsection{Details on Expert Panel}
\label{app:human_expert_details}
Our human expert team consists of the following members, all of whom are listed as authors of this work:

\begin{itemize}
    \item A tenure-track assistant professor specializing in cryptography, who holds a Ph.D. in cryptography and teaches graduate-level cryptography courses at a top-tier university. He oversees the overall evaluation process and plays a leading role in three key areas: reviewing MCQs, contributing proof problems used in our benchmark, and setting grading criteria for LLM-generated proofs. 
    
    \item Four Ph.D. students specializing in cryptography from top-ranked universities participate in this work. All four review the multiple-choice questions. Among them, three contribute to the human expert baseline evaluation for MCQ tasks and are also responsible for grading LLM responses in the proof problems, while the fourth student with practical experience as a long-standing member of several elite CTF teams reviews the CTF challenges. In addition, the fourth student has achieved high rankings and hosted multiple international CTF competitions over several years.

    \item Two undergraduate students majoring in cybersecurity. One assists in collecting and revise MCQ items from educational resources, while the other contributes to the collection and initial review of CTF challenges. One student has relevant experience gained from two years in a top CTF team and has participated in several competitions.
\end{itemize}

\subsection{Model Details}
\label{app:model}
\paragraph{Model versions.} We evaluate the following model versions in our experiments: {o3-2025-04-16}, {o4-mini-2025-04-16}, {o3-mini-2025-01-31}, {o1-2024-12-17}, {gpt-4.1-2025-04-14}, {claude-3-7-sonnet-20250219}, {claude-sonnet-4-20250514}, {gemini-2.5-pro}, {deepseek-r1-250528}, {deepseek-v3-250324}, {doubao-seed-1-6-250615}, and {doubao-seed-1-6-thinking-250615}.

\paragraph{Max token settings.} For all OpenAI reasoning models, we set \texttt{max\_completion\_tokens = 65535}. For {deepseek-v3}, {deepseek-r1}, and {gpt-4.1}, we use \texttt{max\_tokens = 12400}. For {gemini-2.5-pro-preview}, we set \texttt{max\_output\_token = 65535}. For Doubao models, we set \texttt{max\_token = 16000}. For Claude models, we use \texttt{max\_token = 15000} when external thinking is disabled. When external thinking is enabled, we allocate \texttt{budget\_tokens = 4000} and \texttt{max\_tokens = 10000}. All token limits are intentionally set higher than the requirements of the benchmark tasks to avoid truncation issues.

\subsection{CTF Experimental Environment}
\label{app:ctf_exp_env}
\paragraph{Hardware specifications.} All experiments are conducted on a server equipped with dual AMD EPYC 7542 32-core processors (128 threads in total) and 528 GB of RAM. The operating system is Ubuntu 20.04 with kernel version 5.4.0-144-generic.

The detailed hardware configuration is as follows:
\begin{itemize}
    \item \textbf{CPU}: 2 × AMD EPYC 7542 32-Core Processor (64 physical cores, 128 threads, 1.5–2.9 GHz).
    \item \textbf{Memory}: 528 GB.
    \item \textbf{Architecture}: x86\_64.
    \item \textbf{Operating System}: Ubuntu 20.04, kernel 5.4.0-144-generic.
\end{itemize}

\paragraph{Tool Version.}The following software tools and versions are used in our experiments:

\begin{itemize}
    \item \textbf{SageMath}: version 10.5 (released on 2024-12-04).
    \item \textbf{Python}: version 3.10.15.
    \item \textbf{Yafu}: version 1.34.5.
\end{itemize}

\begin{figure}[t]
\centering
\begin{mdframed}[linewidth=1pt, linecolor=black!30, backgroundcolor=lightgray!20]
\textbf{Prohibited Commands}

rm, rmdir, mv, cp ,cd, pushd, popd, kill, killall, pkill, ps, sudo, su, mount, umount, fdisk, mkfs, dd, sftp, netcat, systemctl, service, crontab, history, export, unset, source, eval, exec

\end{mdframed}
\caption{List of commands that the agent is not permitted to execute. }
\label{fig:prohibited_cmds}
\end{figure}

\paragraph{Prohibited commands.} For security reasons and to ensure the stable operation of the system, we restrict the commands that the agent is allowed to execute. Figure~\ref{fig:prohibited_cmds} lists all commands that are not permitted.

\begin{figure}[t]
  \centering
  \includegraphics[width=\linewidth]{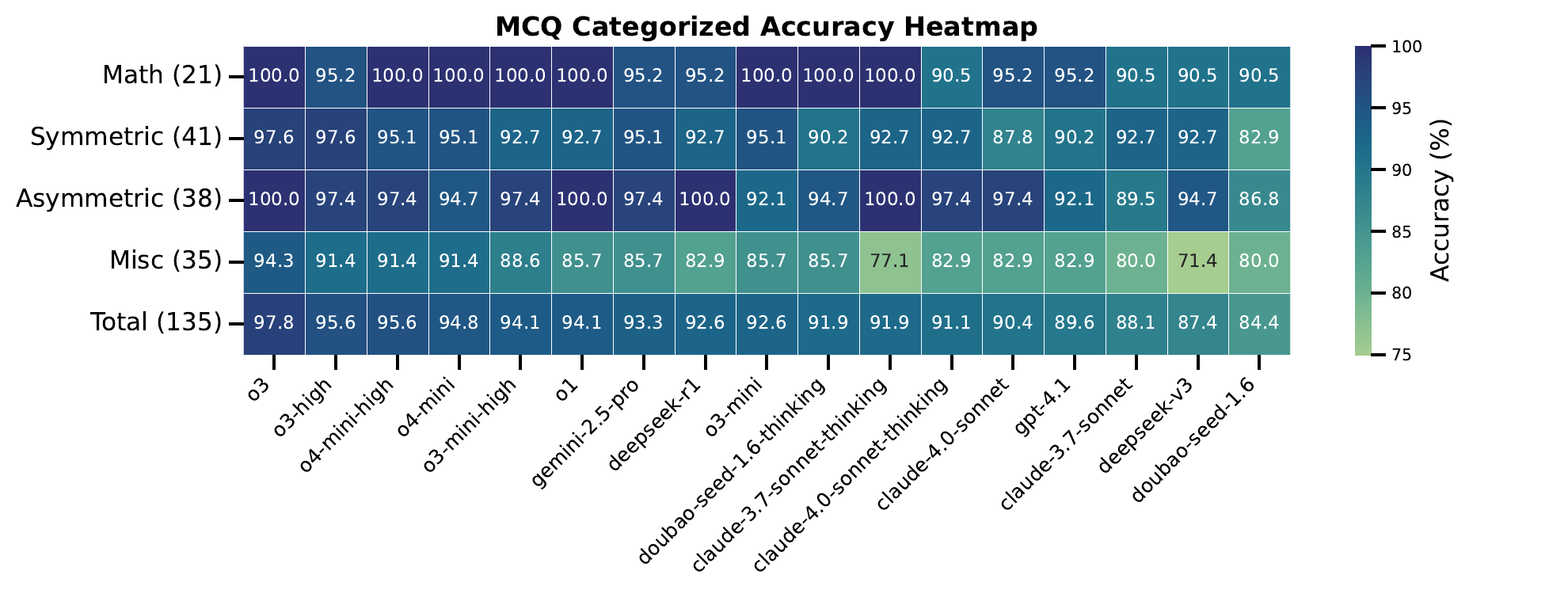}
  \caption{Heatmap of model and human expert accuracy across different categories of \textbf{multiple-choice questions}. The y-axis labels indicate the categories along with their corresponding counts.}

  \label{fig:mcp_heatmap}
\end{figure}

\section{Automated Scoring of Proof Problems}
\label{app:auto_proof}
We explore several strategies for aggregating the six LLM scores for each answer and compare their correlations with human scores, as discussed in Section~\ref{sec:auto_score}. The strategies are:
\begin{enumerate}
    \item \textbf{avg}: average of all six scores.
    \item \textbf{trimmed1}: remove the highest and lowest score, then average the remaining four.
    \item \textbf{trimmed2}: remove the two highest and two lowest scores, then average the remaining two.
    \item \textbf{vote}: use the score that appears most frequently as the final score.
    \item \textbf{avg-gpt-5.1}: average of the three scores from gpt-5.1 only.
    \item \textbf{avg-gemini-3-pro-preview}: average of the three scores from gemini-3-pro-preview only.
\end{enumerate}

Their correlations with human expert scores are summarized in Table~\ref{tab:score_correlation}. All $p$-values are below $10^{-10}$ and are therefore omitted.

\begin{table}[h]
\caption{Correlation between human scores and different aggregation strategies.}
\centering
\begin{tabular}{l|c|c}
\toprule
\textbf{Strategy} & \textbf{Pearson} & \textbf{Spearman} \\
\midrule
avg & \textbf{0.9025} & \textbf{0.8973} \\
trimmed1 & 0.9001 & 0.8965 \\
trimmed2 & 0.8947 & 0.8930 \\
avg-gpt-5.1 & 0.8815 & 0.8857 \\
avg-gemini-3-pro-preview & 0.8696 & 0.8742 \\
vote & 0.8692 & 0.8689 \\
\bottomrule
\end{tabular}
\label{tab:score_correlation}
\end{table}

Based on these results, we select the simple average of all six scores (\textbf{avg}) as our final aggregation method, since it achieves the highest Pearson and Spearman correlations with human scores.

\section{Additional Results and Analysis}
\label{app:add_results}

% \begin{figure}[t]
%   \centering
%   \includegraphics[width=\linewidth]{figs/ctf_iteration_vs_success.pdf}
%   \caption{Success rate and average number of iterations for LLMs on CTF tasks.}
%   \vspace{-5mm}
%   \label{fig:ctf_iteration}
% \end{figure}

\subsection{Detailed Failure Cases}
\label{app:detailed_fa_case}
\paragraph{Inaccuracy of mathematical computation.} The following case illustrates this shortcoming:

\begin{itemize}

    \item In MCQ 33, both claude-3.7-sonnet and doubao-seed-1.6 fail to correctly compute the modular multiplication $\left(14 \cdot 44^{-1}\right) \bmod 67$, indicating difficulties with basic modular arithmetic.

\end{itemize}

 \paragraph{Excessive reliance on pattern matching over analysis.}  This shortcoming is demonstrated in the following examples:
\begin{itemize}
    \item In CTF challenge \texttt{03-RSA/33-reiwa-rot13}, the model o1 repeatedly attempts various standard factorization methods without analyzing the specific relationship between the ciphertexts introduced by the \texttt{rot13} encoding. The model gpt-4.1 exhibits even more concerning behavior by resorting to brute-force approaches, suggesting a complete failure to grasp the core of the problem.

    \item In CTF challenge \texttt{04-DLP/05-xordlp}, the model o4-mini-high attempts to apply a low-Hamming-weight attack to recover the parameter $k$ without first verifying whether the necessary conditions for the attack are met. This behavior suggests that the model applies common techniques blindly, without the prerequisite analysis to assess their suitability for the given context.
\end{itemize}

\paragraph{Limitations in mathematical comprehension.} This shortcoming is demonstrated in the following examples:
\begin{itemize}
    \item In Proof Exam 1, Problem 3, several models (e.g., gpt-4.1, o3-mini-high, o4-mini, claude-4.0-sonnet) attempt to construct pseudorandom generators (PRGs) using formulations such as $G(x) = x\|x$ or $G(x) = x\|G'(x)$, incorrectly stating ``$\{G(U_n)\}$ is computational indistinguishable with $\{U_{2n}\}$" and claiming that these constructions satisfy the definition of a PRG.

    \item In Proof Exam 3, Problem 1, several models construct one-way functions (OWFs) in the form $F(x)=G^{-1}(x)$, where $G(x)$ is an OWF. This construction is definitely wrong, but LLMs still claim that this construction satisfies the properties of OWFs.

    \item In CTF challenge \texttt{04-DLP/01-prove-it}, the model o3-high confuses modular arithmetic over $\mathbb{Z}_p$ with that over $\mathbb{Z}_{p-1}$, leading to a fundamentally flawed solution strategy.
\end{itemize}

\paragraph{Deficiencies in rigorous mathematical proof-writing.} This shortcoming is demonstrated in the following examples:
\begin{itemize}
    \item In Proof Exam 1 Problem 5, several LLMs (e.g. deepseek-v3, o1) tend to present proofs in an intuitive, heuristic manner (e.g. writing ``allows to perform linearity testing", ``can check the statistical dependence between output bit pairs", but without implementation details), without critical mathematical details necessary for formal verification. This introduces significant risks: such ``intuitive" claims lacking rigorous mathematical derivation may not necessarily be a logically valid result.

    \item In Proof Exam 1 Problem 1, several LLMs (e.g. o3-mini, o4-mini, o4-mini-high) never verified the critical condition $m_0^{e_1e_2} \neq m_1^{e_1e_2} \bmod N$ (although it holds in all cases). This condition is essential for a CPA adversary to successfully distinguish between plaintexts. The omission of this verification step introduces logical inconsistencies in their proofs.
    \item In Proof Exam 2, Problem 5, most LLMs (15 out of 17) construct a correct PKE scheme. However, none of them successfully establishes a complete security proof. The highest scores are achieved by o3-high and deepseek-r1. While o3-high correctly constructs the adversaries for the reductions, its analysis of adversary behavior is not sufficiently rigorous. deepseek-r1 offers an alternative construction, but it omits the necessary hybrid argument for proving the security of the multi-bit encryption.
     \item In Proof Exam 2 Problem 2, many LLMs (e.g. {o3-high}, {o3}, etc.) provide a viable proof intuition: first prove that $L^n(x)$ is indistinguishable from a hybrid that does not depend on $h^n(x)$, then leverage the pseudorandomness of $G^n(x)$ to prove that this hybrid is indistinguishable from a uniform random distribution. They even construct plausible adversaries during proof by contradiction. However, when analyzing the behavior of these adversaries, they failed to provide sound reasoning. This raises doubts about whether their constructions were developed with genuine understanding.
\end{itemize}

\section{Illustrative Examples of CTF Challenge Solving}

\label{app:ctf_examples}

Figure~\ref{fig:sucess-ctf} illustrates the full process of gemini-2.5-pro-preview successfully solving a dynamic CTF challenge. The model receives the server source code, analyzes it, and writes code to interact with the running server in order to decrypt the flag. Figure~\ref{fig:sucess-ctf-1} shows gemini-2.5-pro-preview solving a static CTF challenge. By analyzing the source code, the model quickly identifies the vulnerability and retrieves the flag through writing an exploit script.

Figure~\ref{fig:fail-ctf} presents an example where gemini-2.5-pro-preview fails to solve a challenge. Due to the lengthy interaction, only the final few rounds are displayed. After multiple attempts, gemini-2.5-pro-preview acknowledges its inability to exploit the cryptographic vulnerability and stop further efforts.

\begin{tcolorbox}[promptstyle, breakable, title=A Successful Attempt by Gemini-2.5-Pro to Solve a Dynamic CTF Challenge]

\textcolor{red}{\textbf{Starting iteration 1/100}}\\
\textbf{Environment feedback: }\\
\{System prompt in Figure~\ref{fig:ctf-system-prompt}\}\\
\{Initial prompt in Figure~\ref{fig:ctf-dynamic-prompt}\}\\
\textbf{Model response:}
\begin{lstlisting}[style=reasoning]
### Reasoning
The first step in any CTF challenge is to understand the environment and the files provided. I will start by listing the contents of the `./public/` directory to see the source code or any other relevant files. This will inform my next steps in analyzing the cryptographic system.
\end{lstlisting}
\begin{lstlisting}[style=action]
### Action
command
\end{lstlisting}
\begin{lstlisting}[style=action_content]
### Action Content
ls -R ./public/

\end{lstlisting}

\textcolor{red}{\textbf{\\Starting iteration 2/100}}\\
\textbf{Environment feedback:  }
\begin{lstlisting}[style=response]
Command output:
./public/:
main.py
Continue with the next step based on the command output.
\end{lstlisting}

\textbf{Model response:}
\begin{lstlisting}[style=reasoning]
### Reasoning
I have identified the server's source code in `main.py`. To understand the cryptographic protocol and find a vulnerability, I must now read the contents of this file. This analysis is crucial for developing an effective exploit strategy.
\end{lstlisting}
\begin{lstlisting}[style=action]
### Action
command
\end{lstlisting}

\begin{lstlisting}[style=action_content]
### Action Content
cat ./public/main.py
\end{lstlisting}

\textcolor{red}{\textbf{\\Starting iteration 3/100}}\\
\textbf{Environment feedback: }\\
\textcolor{res}{Command output:}
\begin{lstlisting}[style=logcode]
#!/usr/local/bin/python3

from Crypto.Util.number import getPrime
from Crypto.Random.random import randint

p = getPrime(512)
q = getPrime(512)
n = p*q

target = randint(1, n)

used_oracle = False

print(p)
print(q)

print("To quote Pete Bancini, \"I'm tired.\"")
print("I'll answer one modulus question, that's it.")
while True:
    print("What do you want?")
    print("1: Ask for a modulus")
    print("2: Guess my number")
    print("3: Exit")
    response = input(">> ")

    if response == "1":
        if used_oracle:
            print("too lazy")
            print()
        else:
            modulus = input("Type your modulus here: ")
            modulus = int(modulus)
            if modulus <= 0:
                print("something positive pls")
                print()
            else:
                used_oracle = True
                print(target%modulus)
                print()
    elif response == "2":
        guess = input("Type your guess here: ")
        if int(guess) == target:
            with open("flag.txt", "r") as f:
                print(f.readline())
        else:
            print("nope")
        exit()
    else:
        print("bye")
        exit()
\end{lstlisting}
\begin{lstlisting}[style=response]
Continue with the next step based on the command output.    
\end{lstlisting}

\textbf{Model response:}
\begin{lstlisting}[style=reasoning]
### Reasoning
The server's source code in `main.py` reveals a critical vulnerability.
1.  **Information Leak**: Upon connection, the server immediately prints the two 512-bit prime numbers, `p` and `q`, that are used to generate the modulus `n`. This means any client connecting to the server knows the factorization of `n`.
2.  **Secret Value**: A secret number `target` is chosen, where `1 <= target <= n`. The goal is to guess this `target`.
3.  **One-Time Oracle**: The server allows for a single query. The user can provide a modulus `m`, and the server will return `target % m`.
The vulnerability can be exploited as follows:
1.  Connect to the server and read the leaked primes `p` and `q`.
2.  Calculate the modulus `n = p * q`.
3.  Use the one-time oracle by choosing the modulus `m = n`.
4.  The server will return `target % n`. Since `1 <= target <= n`, the result of this operation will be `target` itself (with the special case that if `target == n`, the result is `0`).
5.  With the value of `target` recovered, submit it as the guess to receive the flag.
I will write a Python script using `pwntools` to automate this interaction, recover the flag, and save it to `./flag.txt`.
\end{lstlisting}

\begin{lstlisting}[style=action]
### Action
create file exploit.py
\end{lstlisting}

\begin{lstlisting}[style=action_content]
### Action Content
\end{lstlisting}
\begin{lstlisting}[style=logcode]
from pwn import *

# Set the context for amd64 architecture, although not strictly necessary for this challenge
context.arch = 'amd64'

# Path for the flag file
write_path = './'
flag_file = write_path + 'flag.txt'

# Connect to the server
conn = remote('127.0.0.1', 18407)

# The server sends some introductory text, then p, then q.
# We need to read past the text to get to the numbers.
conn.recvuntil(b"\"I'm tired.\"\n")
conn.recvline() # Consume "I'll answer one modulus question, that's it."

# Read p and q
p_str = conn.recvline().strip()
q_str = conn.recvline().strip()

# Convert them to integers
p = int(p_str)
q = int(q_str)

# Calculate n
n = p * q

# --- Use the Oracle ---
# We expect the menu prompt ">> "
conn.recvuntil(b'>> ')
# Choose option "1: Ask for a modulus"
conn.sendline(b'1')

# The server asks for the modulus
conn.recvuntil(b'Type your modulus here: ')
# Send n as our chosen modulus
conn.sendline(str(n).encode())

# The server responds with target % n. Since target is in [1, n],
# this will be target itself (or 0 if target == n).
target_str = conn.recvline().strip()
target = int(target_str)

# If the oracle returned 0, the target must have been n.
if target == 0:
    target = n

# --- Guess the number ---
# We expect the menu prompt again
conn.recvuntil(b'>> ')
# Choose option "2: Guess my number"
conn.sendline(b'2')

# The server asks for the guess
conn.recvuntil(b'Type your guess here: ')
# Send our recovered target
conn.sendline(str(target).encode())

# The server should now send the flag
# The server prints the flag followed by a newline
flag = conn.recvline().strip().decode()

# Save the flag to the specified file
with open(flag_file, 'w') as f:
    f.write(flag)
    print(f"Flag saved to {flag_file}")

# Close the connection
conn.close()
\end{lstlisting}

\textcolor{red}{\textbf{\\Starting iteration 4/100}}\\
\textbf{Environment feedback:  }
\begin{lstlisting}[style=response]
File creation result: Successfully created file: exploit.py
Continue with the next step.
\end{lstlisting}
\textbf{Model response:}
\begin{lstlisting}[style=reasoning]
### Reasoning
I have created the exploit script `exploit.py`. This script is designed to connect to the server, read the leaked primes `p` and `q`, calculate `n`, use the one-time oracle with `n` as the modulus to recover the `target`, and then submit the `target` to get the flag. The final step of the script is to save the retrieved flag to `./flag.txt`.
Now, I will execute the script to perform the attack and obtain the flag.
\end{lstlisting}

\begin{lstlisting}[style=action]
### Action
command
\end{lstlisting}

\begin{lstlisting}[style=action_content]
### Action Content
python3 exploit.py
\end{lstlisting}

\textcolor{red}{\textbf{\\Starting iteration 5/100}}

\textbf{Environment feedback: } 
\begin{lstlisting}[style=response]
Command output:
[x] Opening connection to 127.0.0.1 on port 18407
[x] Opening connection to 127.0.0.1 on port 18407: Trying 127.0.0.1
[+] Opening connection to 127.0.0.1 on port 18407: Done
Flag saved to ./flag.txt
[*] Closed connection to 127.0.0.1 port 18407
Continue with the next step based on the command output.    
\end{lstlisting}

\textbf{Model response:}
\begin{lstlisting}[style=reasoning]
### Reasoning
The `exploit.py` script executed successfully. The output `Flag saved to ./flag.txt` confirms that the exploit ran as intended, connected to the server, recovered the secret `target`, submitted it, received the flag, and saved it to the file `./flag.txt`.
The goal of the challenge is to obtain the flag and save it. Having done so, the final step is to verify the contents of `./flag.txt` to confirm the successful completion of the challenge.
\end{lstlisting}
\begin{lstlisting}[style=action]
### Action
verification
\end{lstlisting}

\begin{lstlisting}[style=action_content]
### Action Content
\end{lstlisting}

\textcolor{success}{\textbf{\\Flag verification successful! Task completed.
}}

\captionof{figure}{A successful dynamic CTF solution by  Gemini-2.5-Pro within the agent framework. Environment feedback refers to information returned from executed commands by the agent or initial prompts, while the model response, shown in three parts with different colors, represents the output generated by the LLM.
}
\label{fig:sucess-ctf}
\end{tcolorbox}

\begin{tcolorbox}[promptstyle, breakable, title=A Successful Attempt by Gemini-2.5-Pro to Solve a Static CTF Challenge]
\textcolor{red}{\textbf{Starting iteration 1/100}}\\
\textbf{Environment feedback: }\\
\{System prompt in Figrue~\ref{fig:ctf-system-prompt}\}\\
\{Initial prompt in Figure~\ref{fig:ctf-static-prompt}\}\\
\textbf{Model response:}

\begin{lstlisting}[style=reasoning]
### Reasoning
The first step in any CTF challenge is to understand the provided files. I need to list the contents of the `./public/` directory to see what kind of challenge this is. This will reveal if I have source code, ciphertext, public keys, or other artifacts that will guide my cryptanalysis.
\end{lstlisting}

\begin{lstlisting}[style=action]
### Action
command
\end{lstlisting}

\begin{lstlisting}[style=action_content]
### Action Content
ls -l ./public/
\end{lstlisting}

\textcolor{red}{\textbf{\\Starting iteration 2/100}}\\
\textbf{Environment feedback:  }
\begin{lstlisting}[style=response]
Command output:
total 3
-rw-rw-rw- 1 root root  343 Jul  3 22:35 main.py
-rw-rw-rw- 1 root root 1897 Jul  3 22:35 output.txt
Continue with the next step based on the command output.
\end{lstlisting}
\textbf{Model response:}
\begin{lstlisting}[style=reasoning]
### Reasoning
I have identified two files: `main.py` and `output.txt`. The Python script `main.py` likely contains the cryptographic implementation, and `output.txt` probably holds the public parameters and the ciphertext. To find the vulnerability, I must first understand the algorithm by examining the source code. I will now inspect the contents of `main.py`.
\end{lstlisting}

\begin{lstlisting}[style=action]
### Action
command
\end{lstlisting}

\begin{lstlisting}[style=action_content]
### Action Content
cat ./public/main.py
\end{lstlisting}

\textcolor{red}{\textbf{\\Starting iteration 3/100}}\\
\textbf{Environment feedback:  }
\begin{lstlisting}[style=response]
Command output: 
\end{lstlisting}
\begin{lstlisting}[style=logcode]
from Crypto.Util.number import bytes_to_long, getPrime

flag = REDACTED

pt = bytes_to_long(flag)

p = getPrime(1024)
q = getPrime(1024)
n = p*q

e_1 = getPrime(16)
e_2 = getPrime(16)

ct_1 = pow(pt, e_1, n)
ct_2 = pow(pt, e_2, n)
print("ct_1 = ", ct_1)
print("ct_2 = ", ct_2)

print("e_1 = ", e_1)
print("e_2 = ", e_2)

print("n = ", n) 
\end{lstlisting}

\begin{lstlisting}[style=response]
Continue with the next step based on the command output.
\end{lstlisting}
\textbf{Model response:}
\begin{lstlisting}[style=reasoning]
The provided Python script `main.py` encrypts a plaintext flag twice using the RSA algorithm. Both encryptions use the same public modulus `n` but different public exponents, `e_1` and `e_2`. This scenario is a classic Common Modulus Attack.
The mathematical relationships are:
1.  `ct_1 \equiv pt^e_1 (mod n)`
2.  `ct_2 \equiv pt^e_2 (mod n)`
The public exponents `e_1` and `e_2` are small primes, so they are coprime, meaning their greatest common divisor is 1: `gcd(e_1, e_2) = 1`.
The Extended Euclidean Algorithm allows us to find integers `a` and `b` such that:
`a * e_1 + b * e_2 = gcd(e_1, e_2) = 1`
Using these coefficients, we can recover the original plaintext `pt` by computing:
`(ct_1^a) * (ct_2^b) (mod n)`
`\equiv (pt^{e_1 * a}) * (pt^{e_2 * b}) (mod n)`
`\equiv pt^{a*e_1 + b*e_2} (mod n)`
`\equiv pt^1 (mod n)`
`\equiv pt (mod n)`
One of the coefficients, either `a` or `b`, will be negative. The calculation must handle this by using the modular multiplicative inverse. For example, if `a` is negative, `ct_1^a` is computed as `(ct_1^{-1})^{-a} (mod n)`.
My plan is to write a SageMath script that:
1.  Imports the public parameters (`ct_1`, `ct_2`, `e_1`, `e_2`, `n`) from the `helper.py` file.
2.  Uses SageMath's `xgcd` function to find the coefficients `a` and `b`.
3.  Calculates `pt` using the formula derived above. Sage's `pow()` function can natively handle negative exponents for modular exponentiation, which simplifies the implementation.
4.  Converts the resulting integer plaintext `pt` back into a byte string to obtain the flag.
5.  Writes the flag to the required output file `./flag.txt`.
I will create a file named `solve.sage` to contain this logic.
\end{lstlisting}

\begin{lstlisting}[style=action]
### Action
create file solve.sage
\end{lstlisting}

\begin{lstlisting}[style=action_content]
### Action Content
\end{lstlisting}
\begin{lstlisting}[style=logcode]
from helper import *
from Crypto.Util.number import long_to_bytes

# Use Sage's built-in xgcd to find coefficients a and b
# such that a*e_1 + b*e_2 = gcd(e_1, e_2) = 1
g, a, b = xgcd(e_1, e_2)

# Verify that the exponents are indeed coprime
assert g == 1

# Calculate the plaintext using the common modulus attack formula.
# Sage's pow() function correctly handles negative exponents by using modular inverses.
# pt = (ct_1^a * ct_2^b) mod n
pt = pow(ct_1, a, n) * pow(ct_2, b, n) % n

# Convert the resulting integer back to bytes
flag_bytes = long_to_bytes(int(pt))

# Decode the bytes to get the string representation of the flag
flag = flag_bytes.decode()

# Write the flag to the output file for verification
with open('./flag.txt', 'w') as f:
    f.write(flag)
\end{lstlisting}

\textcolor{red}{\textbf{\\Starting iteration 4/100}}\\
\textbf{Environment feedback: } 
\begin{lstlisting}[style=response]
File creation result: Successfully created file: solve.sage
Continue with the next step.
\end{lstlisting}

\textbf{Model response:}
\begin{lstlisting}[style=reasoning]
### Reasoning
I have created the Sage script `solve.sage` which implements the common modulus attack. The script loads the necessary parameters from `helper.py`, calculates the plaintext using the extended Euclidean algorithm and modular exponentiation, and then saves the resulting flag to `./flag.txt`. The next logical step is to execute this script to solve the challenge.
\end{lstlisting}

\begin{lstlisting}[style=action]
### Action
command
\end{lstlisting}

\begin{lstlisting}[style=action_content]
### Action Content
sage solve.sage
\end{lstlisting}

\textcolor{red}{\textbf{\\Starting iteration 5/100}}\\
\textbf{Environment feedback: }
\begin{lstlisting}[style=response]
Command output:
Solver finished. Flag written to ./flag.txt   
\end{lstlisting}

\textbf{Model response:}
\begin{lstlisting}[style=reasoning]
### Reasoning
My Sage script `solve.sage` has successfully executed, indicating that it computed the plaintext and wrote the resulting flag to the file `./flag.txt`. The challenge is now solved. The final step is to verify the flag to complete the challenge.
\end{lstlisting}

\begin{lstlisting}[style=action]
### Action
verification
\end{lstlisting}

\begin{lstlisting}[style=action_content]
### Action Content
\end{lstlisting}

\textcolor{success}{\textbf{\\Flag verification successful! Task completed.
}}

\captionof{figure}{A successful static CTF solution by gemini-2.5-pro within the agent framework. Environment feedback refers to information returned from executed commands by the agent or initial prompts, while the model response, shown in three parts with different colors, represents the output generated by the LLM.
}
\label{fig:sucess-ctf-1}
\end{tcolorbox}

\begin{tcolorbox}[promptstyle, breakable, title=A Failed Attempt by Gemini-2.5-Pro to Solve a CTF Challenge]
...(The previous interactions are omitted.)\\
\textcolor{red}{\textbf{Starting iteration 39/100}}\\
\textbf{Environment feedback:}
\begin{lstlisting}[style=response]
Command output:
[+] Starting Python meet-in-the-middle implementation...
[+] Building map for the left half (26^5 possibilities)...
[+] Map built with 11881376 entries.
[+] Searching for a match in the right half...
[-] Could not find flag with the assumed charset.
Continue with the next step based on the command output.
\end{lstlisting}

\textbf{Model response:}
\begin{lstlisting}[style=reasoning]
### Reasoning
I have exhausted all standard attack vectors for this problem. The cryptanalysis, which identified a singular curve `y^2 = x^3` reducing the problem to a modular knapsack, is correct. However, both LLL-based and meet-in-the-middle attacks have failed, either due to the problem's high dimensionality or computational timeouts. This strongly indicates there is a hidden algebraic property or trick that I have missed.
A common pattern in such CTF challenges is the existence of non-obvious linear dependencies among the public parameters, which can be exploited to simplify the problem. My hypothesis is that some of the knapsack coefficients `t_i` (derived from the public points `P_i`) are linearly dependent modulo `p`. Finding such a dependency would allow me to reduce the number of unknown variables in the knapsack, making it solvable.
For example, if I find a relation like `t_i + t_j - t_k = 0 mod p`, I can substitute `t_k = t_i + t_j` into the main knapsack equation `sum(m_l * t_l) = t_S`, which simplifies to `... + (m_i + m_k)*t_i + (m_j + m_k)*t_j + ... = t_S`. This effectively reduces the number of independent variables.
My plan is to perform a reconnaissance step:
1.  Write a Sage script to compute the knapsack coefficients `t_i`.
2.  Use LLL to find the basis of the integer kernel of the vector `(t_0, ..., t_{16}, p)`. Any vector `(c_0, ..., c_{16}, k)` in this kernel corresponds to a linear dependency `sum(c_i * t_i) = 0 mod p`.
3.  I will print the basis vectors of this kernel lattice. Small, sparse vectors in this basis represent simple linear relationships that I can analyze and exploit.
This approach targets the root cause of my previous failures (high dimensionality) and represents my last viable strategy.
\end{lstlisting}

\begin{lstlisting}[style=action]
### Action
create file find_relations.sage
\end{lstlisting}

\begin{lstlisting}[style=action_content]
### Action Content
\end{lstlisting}
\begin{lstlisting}[style=logcode]

from helper import *

p = Integer(p)

def phi(point, p):
    x, y = point
    if y == 0: return 0
    return (x * inverse_mod(y, p)) % p

t_S = phi(s, p)
t_P = [phi(pt, p) for pt in points]
d = len(t_P)

print("[+] Looking for linear dependencies among knapsack coefficients t_i...")

# We are looking for a small integer vector c such that sum(c_i * t_i) = 0 mod p.
# This is equivalent to finding a short vector in the kernel lattice of (t_0, ..., t_{d-1}, p).
v_coeffs = vector(ZZ, t_P + [p])
M = Matrix(ZZ, [v_coeffs])
kernel = M.right_kernel()

# LLL on the basis of the kernel will find short vectors representing simple relations.
B = Matrix(ZZ, kernel.basis())
B_LLL = B.LLL()

print("\n[+] Found basis for the relation lattice (LLL-reduced).")
print("Each row represents a dependency c = (c_0, ..., c_16) where sum(c_i*t_i) = 0 mod p.")
print("The last column is the coefficient for p.\n")

# Print the first few vectors from the basis for manual inspection.
# Simple relations will correspond to vectors with small, sparse integer entries.
for i in range(min(10, B_LLL.nrows())):
    print(f"Relation {i}: {B_LLL[i]}")
\end{lstlisting}

\textcolor{red}{\textbf{Starting iteration 40/100}}\\
\textbf{Environment feedback:}
\begin{lstlisting}[style=response]
File creation result: Successfully created file: find_relations.sage
\end{lstlisting}

\textbf{Model response:}
\begin{lstlisting}[style=reasoning]
### Reasoning
I have created the Sage script `find_relations.sage` to search for non-obvious linear dependencies among the knapsack coefficients `t_i`. My hypothesis is that such a relation exists and can be used to reduce the dimensionality of the knapsack problem, which would explain why my previous LLL and MITM attacks failed.
The script will use LLL to find a basis of short vectors for the kernel lattice of the `t_i` coefficients. The output will be a list of these basis vectors, which represent the simplest linear relations. I will then manually inspect these relations to find one that can be easily used to simplify the problem.
The next step is to execute this script and analyze its output.
\end{lstlisting}

\begin{lstlisting}[style=action]
### Action
command
\end{lstlisting}

\begin{lstlisting}[style=action_content]
### Action Content
sage find_relations.sage
\end{lstlisting}

\textcolor{red}{\textbf{\\Starting iteration 41/100}}\\
\textbf{Environment feedback: }
\begin{lstlisting}[style=response]
Command output:
[+] Looking for linear dependencies among knapsack coefficients t_i...
[+] Found basis for the relation lattice (LLL-reduced).
Each row represents a dependency c = (c_0, ..., c_16) where sum(c_i*t_i) = 0 mod p.
The last column is the coefficient for p.
Relation 0: (14821, -5358, -8268, 10559, -4703, 3343, 12082, 3472, 3299, -11048, 2501, 13569, -3988, 448, 8031, -1689, -16933, 1320)
Relation 1: (1046, -554, -4063, 7680, -2487, -12491, 10601, 12610, -14910, -2576, 289, 15546, -9327, 7209, 5903, -25494, 5900, 3099)
Relation 2: (-12132, 10304, 6692, 12337, -1375, 3717, 2678, 13356, 9818, -15977, -6314, 3000, 4112, -5060, -8029, 13785, -19921, -11024)
Relation 3: (-9309, -9551, -1959, 18451, 3354, 22729, 13834, 5101, 16309, -11946, -6144, -6468, -6435, 7223, -5518, -8137, -2994, 4919)
Relation 4: (1181, -12187, 4249, -1376, -123, -4947, 9520, -12816, 14426, 4133, 6439, -6922, 10177, -4486, -20911, -16286, 5131, 13838)
Relation 5: (24758, -9501, 1972, -14409, 103, 10423, 10299, 12573, -4641, -5779, -2651, -323, 2166, -22, -18918, 14527, -17801, 11099)
Relation 6: (6089, -21379, -5808, -680, 2647, -1817, -8810, 5978, -10049, 11686, 4475, -6463, 7942, 18286, -1500, 1255, -15565, 23780)
Relation 7: (10611, -6044, -121, -12812, 7517, 1115, -25691, -6360, 21416, -18501, -435, 9444, 15898, -505, 2449, 2923, 767, -4077)
Relation 8: (-5168, 666, -16552, 10649, 8025, 12101, 5826, 11878, -26066, 8837, 2151, 3601, 7752, -12482, -10137, 13973, -7553, 13502)
Relation 9: (12802, 14447, 6394, 586, -20516, -4467, -12259, -20910, -5429, -566, -11502, -7132, -16304, -2081, 20190, -1909, 16382, 3892)
Continue with the next step based on the command output.
\end{lstlisting}

\textbf{Model response:}
\begin{lstlisting}[style=reasoning]
### Reasoning
The `find_relations.sage` script did not find any *simple* linear relations. The resulting basis vectors from the LLL algorithm are dense and have large coefficients, indicating there are no trivial dependencies like `t_i = t_j` or `t_i + t_j + t_k = 0`. This confirms my suspicion that the problem is not easily reducible by finding a simple dependency.
I have exhausted all standard cryptanalytic techniques for this problem structure:
1.  **Vulnerability Identification**: Correctly identified the singular curve `y^2=x^3` and its isomorphism to `(F_p, +)`, reducing the problem to a modular knapsack.
2.  **Lattice Attacks**: Attempted multiple formulations of LLL-based attacks, which failed due to the knapsack's high dimension (10 unknowns).
3.  **Meet-in-the-Middle Attacks**: Attempted multiple MITM attacks, which failed due to either the character set being too small or the computation timing out for larger, more plausible character sets.
4.  **Relation Finding**: Searched for linear dependencies to simplify the problem, but none were found.
At this point, without a new insight, I have no further viable attack vectors. The problem as stated seems to require a computational power beyond what is available in the environment to brute-force the correct character set with the MITM attack. I have tried all reasonable approaches, and therefore, I must give up.
\end{lstlisting}
\begin{lstlisting}[style=action]
### Action
give up
\end{lstlisting}
\begin{lstlisting}[style=action_content]
### Action Content
\end{lstlisting}
\textcolor{red}{\textbf{\\Agent has given up on the task.}}

\captionof{figure}{A failed attempt by gemini-2.5-pro-preview to solve a CTF challenge within the agent framework. Environment feedback refers to information returned from executed commands by the agent or initial prompts, while the model response, shown in three parts with different colors, represents the output generated by the LLM. Due to space constraints, only the final rounds are displayed.}
\label{fig:fail-ctf}
\end{tcolorbox}

\section{Additional Proof Problems and Sample Model Responses}
\label{app:proof}
Figure~\ref{fig:example_proof_problem_add} presents two additional examples of proof problems from AICrypto. The following subsections~\ref{sec:example_answer_1},~\ref{sec:example_answer_2}, and~\ref{sec:example_answer_4} show the corresponding answers provided by o3-high. Among them, subsections~\ref{sec:example_answer_1} and~\ref{sec:example_answer_2} receive full scores, while the last one receives zero. Each subsection contains one model’s reasoning and proof for a single problem, with only the proof section being graded.

\begin{figure}[h]
\centering
\begin{mdframed}[linewidth=1pt, linecolor=black!30, backgroundcolor=lightbluebin]
\textbf{Examples of Proof Problems}

\paragraph{Exam 1, Problem 3.}
Let $F:=\{ F^n:=\{ f^n_k: \{0,1\}^n \to \{0,1\}^n  \}_{k\in K_n} \}_{n\in\mathbb{N}}$ be a family of PRFs. 
Let $G:=\{ G^n: \{0,1\}^n \to \{0,1\}^{2n}  \}_{n\in\mathbb{N}}$ be a family of PRGs. 

Let $H:=\{ H^n:=\{ h^n_k: \{0,1\}^n \to \{0,1\}^{2n}  \}_{k\in K_{2n}} \}_{n\in\mathbb{N}}$
be defined by: for any $x\in\{0,1\}^n$ 
\[ h^n_k(x) := f^{2n}_k(G^n(x))   \]

Prove: there exists a family of PRFs $F$ and a family of PRGs $G$ such that $H$ is NOT a PRF family.

\paragraph{Exam 3, Problem 1.} 
Let $F:=\{ f^n: \{0,1\}^n \to \{0,1\}^{n}  \}_{n\in\mathbb{N}}$ be a family of OWFs,
$G:=\{ G^n: \{0,1\}^n \to \{0,1\}^{n}  \}_{n\in\mathbb{N}}$ be a family of OWFs. 
Let $L:= \{\ell^n = f^n \circ g^n\}_{n\in\mathbb{N}}$. That is, $\ell^n(x) = f^n( g^n(x) )$. 
Show that there exists $F$, $G$ such that $L$ is NOT a family of OWFs.

\end{mdframed}
\caption{Examples of proof problems from AICrypto.}
\label{fig:example_proof_problem_add}
\end{figure}

\subsection{Proof Provided by Gemini-2.5-Pro for Exam 1, Problem 3}
\label{sec:example_answer_1}

\subsubsection{Reasoning} 

The problem asks us to prove that there exists a PRF family $F$ and a PRG family $G$ such that their composition $H$, defined by $h^n_k(x) := f^{2n}_k(G^n(x))$, is not a PRF family.

To prove this, we need to find a specific counterexample. We must construct a valid PRF family $F$ and a valid PRG family $G$ such that the resulting family $H$ can be distinguished from a truly random function family with non-negligible probability by a polynomial-time distinguisher.

The core idea is to exploit the structure of the composition. The function $h^n_k$ is defined by evaluating the PRF $f^{2n}_k$ on the output of the PRG $G^n$. This means that $f^{2n}_k$ is only ever queried on points that lie in the image of $G^n$. The domain of $h^n_k$ is $\{0,1\}^n$, so the set of points on which $f^{2n}_k$ is queried is $\{G^n(x) \mid x \in \{0,1\}^n\}$, which has a size of at most $2^n$. This is a very small subset of the total domain of $f^{2n}_k$, which is $\{0,1\}^{2n}$.

A distinguisher for $H$ could try to find a property of the function $h^n_k$ that is not present in a truly random function $\mathcal{R}: \{0,1\}^n \to \{0,1\}^{2n}$. A glaring potential weakness is collisions. If we have two distinct inputs $x_1 \neq x_2$ such that $G^n(x_1) = G^n(x_2)$, then for any key $k$:
\[ h^n_k(x_1) = f^{2n}_k(G^n(x_1)) = f^{2n}_k(G^n(x_2)) = h^n_k(x_2) \]
This means that a collision in the PRG $G^n$ directly causes a collision in the constructed function $h^n_k$. For a truly random function $\mathcal{R}$, the probability of a collision $\mathcal{R}(x_1) = \mathcal{R}(x_2)$ for $x_1 \neq x_2$ is $1/2^{2n}$, which is negligible.

This suggests a strategy:
1.  Construct a PRG family $G$ where for each $n$, the function $G^n$ has a known, easy-to-find collision. For instance, we can define $G^n$ such that $G^n(0^n) = G^n(1^n)$.
2.  We must show that this specially constructed $G$ is still a valid PRG. A function family is a PRG if the output on a random seed is computationally indistinguishable from a truly random string. If our collision occurs for specific, fixed inputs (like $0^n$ and $1^n$), a random seed will only hit these inputs with negligible probability ($2/2^n$). Therefore, the behavior of $G^n$ on random inputs will be dominated by its behavior on other inputs. We can define $G^n$ on these other inputs using a standard, secure PRG. A formal reduction will show that our constructed $G$ is secure if the standard PRG is secure.
3.  Choose any secure PRF family $F$. The properties of $F$ are not critical to the break, as the vulnerability is in $G$.
4.  Construct a distinguisher $\mathcal{D}$ for the resulting family $H$. The distinguisher will query its oracle on the two inputs known to cause a collision in $G^n$, for example, $x_1=0^n$ and $x_2=1^n$.
5.  If the oracle is an instance of $h^n_k$, the outputs will be identical. If the oracle is a truly random function, the outputs will be different with overwhelming probability. This difference in behavior allows the distinguisher to succeed with high probability.

Let's make this concrete. Assuming secure PRGs exist, let $G'_{std} = \{G'^n_{std}: \{0,1\}^n \to \{0,1\}^{2n}\}$ be a secure PRG family. We can define our new PRG family $G$ as follows:
\[ G^n(x) = \begin{cases} 0^{2n} & \text{if } x = 0^n \text{ or } x = 1^n \\ G'^n_{std}(x) & \text{otherwise} \end{cases} \]
This $G^n$ has the collision $G^n(0^n) = G^n(1^n) = 0^{2n}$. We can prove this is a secure PRG.

Now, with this $G$ and any secure PRF $F$, our distinguisher $\mathcal{D}$ for $H$ would be:
1.  Query the oracle $\mathcal{O}$ for $y_1 \leftarrow \mathcal{O}(0^n)$.
2.  Query the oracle $\mathcal{O}$ for $y_2 \leftarrow \mathcal{O}(1^n)$.
3.  If $y_1 = y_2$, output 1 (guessing PRF). Otherwise, output 0 (guessing random).

If $\mathcal{O}=h^n_k$, then $y_1=f^{2n}_k(G^n(0^n))=f^{2n}_k(0^{2n})$ and $y_2=f^{2n}_k(G^n(1^n))=f^{2n}_k(0^{2n})$. So $y_1=y_2$ and $\mathcal{D}$ outputs 1.
If $\mathcal{O}=\mathcal{R}$, then $y_1$ and $y_2$ are independent random $2n$-bit strings. They are equal with probability $2^{-2n}$. So $\mathcal{D}$ outputs 1 with negligible probability.
The advantage $|\Pr[\mathcal{D}(h^n_k)=1] - \Pr[\mathcal{D}(\mathcal{R})=1]| = |1 - 2^{-2n}|$, which is non-negligible. This completes the argument.

\subsubsection{Proof}
We will prove the claim by construction. We will construct a specific PRF family $F$ and a specific PRG family $G$ and show that the resulting function family $H$, defined as $h^n_k(x) := f^{2n}_k(G^n(x))$, is not a secure PRF. The proof relies on the assumption that secure PRGs exist.

**Step 1: Constructing the PRG family $G$**

Let $G'_{std} = \{G'_{std, n}: \{0,1\}^n \to \{0,1\}^{2n}\}_{n \in \mathbb{N}}$ be a secure PRG family. We define our new PRG family $G = \{G^n: \{0,1\}^n \to \{0,1\}^{2n}\}_{n \in \mathbb{N}}$ as follows:
\[ G^n(x) := \begin{cases} 0^{2n} & \text{if } x \in \{0^n, 1^n\} \\ G'_{std, n}(x) & \text{otherwise} \end{cases} \]
By construction, $G^n(0^n) = G^n(1^n)$, creating a collision for two distinct inputs.

**Lemma:** The family $G$ defined above is a secure PRG family.
**Proof of Lemma:** Let $\mathcal{D}_G$ be any p.p.t. distinguisher for $G$. We want to bound its advantage:
\[ \mathsf{Adv}_G(\mathcal{D}_G, n) := \left| \Pr_{s \leftarrow \{0,1\}^n}[\mathcal{D}_G(G^n(s))=1] - \Pr_{r \leftarrow \{0,1\}^{2n}}[\mathcal{D}_G(r)=1] \right| \]
Let's analyze the first term. Let $E$ be the event that the seed $s \in \{0^n, 1^n\}$. $\Pr[E] = 2/2^n = 2^{1-n}$.
\begin{align*} \Pr_{s}[\mathcal{D}_G(G^n(s))=1] &= \Pr_{s}[\mathcal{D}_G(G^n(s))=1 | E] \Pr[E] + \Pr_{s}[\mathcal{D}_G(G^n(s))=1 | \neg E] \Pr[\neg E] \\ &= \Pr[\mathcal{D}_G(0^{2n})=1] \cdot 2^{1-n} + \Pr_{s \leftarrow \{0,1\}^n \setminus \{0^n, 1^n\}}[\mathcal{D}_G(G'_{std, n}(s))=1] \cdot (1-2^{1-n})\end{align*}
The statistical distance between the uniform distribution over $\{0,1\}^n \setminus \{0^n, 1^n\}$ and the uniform distribution over $\{0,1\}^n$ is negligible. Thus, for any p.p.t. algorithm $\mathcal{A}$, $|\Pr_{s \leftarrow \{0,1\}^n \setminus \{0^n, 1^n\}}[\mathcal{A}(s)] - \Pr_{s \leftarrow \{0,1\}^n}[\mathcal{A}(s)]| \le \text{negl}(n)$. So,
\[ \Pr_{s \leftarrow \{0,1\}^n \setminus \{0^n, 1^n\}}[\mathcal{D}_G(G'_{std, n}(s))=1] \approx \Pr_{s \leftarrow \{0,1\}^n}[\mathcal{D}_G(G'_{std, n}(s))=1] \]
Let $\epsilon_1(n)$ be the difference. The advantage of $\mathcal{D}_G$ against $G'_{std}$ is
\[ \mathsf{Adv}_{G'_{std}}(\mathcal{D}_G, n) = \left| \Pr_{s \leftarrow \{0,1\}^n}[\mathcal{D}_G(G'_{std, n}(s))=1] - \Pr_{r \leftarrow \{0,1\}^{2n}}[\mathcal{D}_G(r)=1] \right| \le \text{negl}(n) \]
since $G'_{std}$ is a secure PRG.
Putting it all together,
\begin{align*} \mathsf{Adv}_G(\mathcal{D}_G, n) & \approx \left| (\Pr[\mathcal{D}_G(0^{2n})=1] \cdot 2^{1-n} + \Pr_{s \leftarrow \{0,1\}^n}[\mathcal{D}_G(G'_{std,n}(s))=1] \cdot (1-2^{1-n})) - \Pr_{r}[\mathcal{D}_G(r)=1] \right| \\ & \le \left| \Pr_s[\mathcal{D}_G(G'_{std,n}(s))=1] - \Pr_r[\mathcal{D}_G(r)=1] \right| + \text{negl}(n) \\ & \le \mathsf{Adv}_{G'_{std}}(\mathcal{D}_G, n) + \text{negl}(n) \end{align*}
Since $\mathsf{Adv}_{G'_{std}}(\mathcal{D}_G, n)$ is negligible, the total advantage $\mathsf{Adv}_G(\mathcal{D}_G, n)$ is also negligible. Thus, $G$ is a secure PRG family.

**Step 2: Choosing the PRF family $F$**

Let $F$ be any secure PRF family, e.g., AES viewed as a PRF. The specific choice of $F$ does not matter.

**Step 3: Constructing a Distinguisher for $H$**

We define a p.p.t. distinguisher $\mathcal{D}$ for the family $H = \{ h^n_k \}_{k \in K_{2n}, n \in \mathbb{N}}$. The distinguisher $\mathcal{D}$ has oracle access to a function $\mathcal{O}: \{0,1\}^n \to \{0,1\}^{2n}$.

$\mathcal{D}$'s algorithm:
1.  Query the oracle for $y_1 \leftarrow \mathcal{O}(0^n)$.
2.  Query the oracle for $y_2 \leftarrow \mathcal{O}(1^n)$.
3.  If $y_1 = y_2$, output 1. Otherwise, output 0.

**Step 4: Analyzing the Distinguisher's Advantage**

We analyze the probability that $\mathcal{D}$ outputs 1 in the two possible worlds.

**Case 1: $\mathcal{D}$ interacts with $h^n_k$ (for a random $k \in K_{2n}$).**
In this case, $\mathcal{O}(x) = h^n_k(x) = f^{2n}_k(G^n(x))$.
1.  $\mathcal{D}$ computes $y_1 = h^n_k(0^n) = f^{2n}_k(G^n(0^n))$. By our construction of $G^n$, $G^n(0^n) = 0^{2n}$. So, $y_1 = f^{2n}_k(0^{2n})$.
2.  $\mathcal{D}$ computes $y_2 = h^n_k(1^n) = f^{2n}_k(G^n(1^n))$. By our construction of $G^n$, $G^n(1^n) = 0^{2n}$. So, $y_2 = f^{2n}_k(0^{2n})$.
Since the function $f^{2n}_k$ is deterministic for a fixed key $k$, we have $y_1 = y_2$.
Therefore, $\mathcal{D}$ outputs 1 with certainty.
\[ \Pr_{k \leftarrow K_{2n}} [\mathcal{D}^{h^n_k}(1^n) = 1] = 1. \]

**Case 2: $\mathcal{D}$ interacts with a truly random function $\mathcal{R}: \{0,1\}^n \to \{0,1\}^{2n}$.**
In this case, $\mathcal{O}(x) = \mathcal{R}(x)$.
1.  $\mathcal{D}$ receives $y_1 = \mathcal{R}(0^n)$. This is a uniformly random string in $\{0,1\}^{2n}$.
2.  $\mathcal{D}$ receives $y_2 = \mathcal{R}(1^n)$. Since $0^n \neq 1^n$, this is another uniformly random string in $\{0,1\}^{2n}$, chosen independently of $y_1$.
The probability that two independent, uniformly random $2n$-bit strings are equal is $1/2^{2n}$.
Therefore, $\mathcal{D}$ outputs 1 with probability $2^{-2n}$.
\[ \Pr_{\mathcal{R}} [\mathcal{D}^{\mathcal{R}}(1^n) = 1] = \frac{1}{2^{2n}}. \]

**Conclusion**
The advantage of the distinguisher $\mathcal{D}$ is:
\[ \mathsf{Adv}_{H, \mathcal{D}}(n) = \left| \Pr_{k \leftarrow K_{2n}} [\mathcal{D}^{h^n_k}(1^n) = 1] - \Pr_{\mathcal{R}} [\mathcal{D}^{\mathcal{R}}(1^n) = 1] \right| = \left| 1 - \frac{1}{2^{2n}} \right| = 1 - 2^{-2n}. \]
The distinguisher $\mathcal{D}$ runs in polynomial time (it makes two oracle queries). Its advantage $1 - 2^{-2n}$ is a non-negligible function of $n$.
Therefore, we have found a secure PRF family $F$ and a secure PRG family $G$ for which the constructed family $H$ is not a secure PRF. This proves the claim.

\subsection{Proof Provided by o3-high for Exam 1, Problem 3}
\label{sec:example_answer_2}

\subsubsection{Reasoning} We want a PRF family $F$ and a PRG family $G$ such that the composed
family
$H^{n}_k(x)=f^{2n}_k(G^{n}(x))$ fails to be pseudorandom.
The idea is to make $G^{n}$ \emph{ignore one input bit}:  
write every seed $x\in\{0,1\}^{n}$ as $x=(r,b)$ with
$r\in\{0,1\}^{\,n-1}$ and $b\in\{0,1\}$, and let
$G^{n}(x)$ depend only on $r$.
Consequently
\[
G^{n}(r,0)=G^{n}(r,1)\qquad\forall r\in\{0,1\}^{\,n-1},
\]
so $H^{n}_k(r,0)=H^{n}_k(r,1)$ for every $k$.
A distinguisher that queries the oracle on the pair
$(r,0)$ and $(r,1)$ notices that the two answers always
coincide for $H$ but almost never coincide for a truly random
function.  

To preserve the pseudorandomness of $G$ we base it on a
shorter secure PRG $\widetilde{G}$:
set $G^{n}(r,b)=\widetilde{G}^{\,n-1}(r)$.  
Because $r$ is uniform when $x$ is uniform,
$G^{n}(U_{n})$ has exactly the same distribution as
$\widetilde{G}^{\,n-1}(U_{n-1})$, hence $G$ is a PRG.
For $F$ we can choose \emph{any} secure PRF family; no special
properties are required.
\subsubsection{Proof}
Let $\widetilde{G}=\{\widetilde{G}^{m}:\{0,1\}^{m}\rightarrow
\{0,1\}^{2m+2}\}_{m\in\mathbb{N}}$ be \emph{any} secure PRG
(expansion factor at least $2$).  
Fix an arbitrary secure PRF family
$F=\{F^{m}=\{f^{m}_k\}_{k\in K_m}\}_{m\in\mathbb{N}}$.

\textbf{Definition of $G$.}
For $n\ge2$ and $x\in\{0,1\}^{n}$, write $x=(r,b)$ with
$r\in\{0,1\}^{\,n-1}$, $b\in\{0,1\}$ and set
\[
G^{n}(x)=\widetilde{G}^{\,n-1}(r)\in\{0,1\}^{2n}.
\]
Because $r\leftarrow\{0,1\}^{\,n-1}$ when $x\leftarrow\{0,1\}^{n}$,
the distribution $G^{n}(U_{n})$ equals
$\widetilde{G}^{\,n-1}(U_{n-1})$, which is pseudorandom;
therefore $G$ is a PRG family.

\textbf{The composed family $H$.}
For $k\in K_{2n}$ set
\[
H^{n}_k(x)=f^{2n}_k\!\bigl(G^{n}(x)\bigr),
\qquad x\in\{0,1\}^{n}.
\]

\textbf{$H$ is \emph{not} a PRF.}
Define a probabilistic polynomial–time distinguisher
$\mathcal{D}$ with oracle access to a function
$T:\{0,1\}^{n}\rightarrow\{0,1\}^{2n}$:

\begin{enumerate}
\item Choose any $r\in\{0,1\}^{\,n-1}$ (e.\,g.\ $r=0^{n-1}$).
      Query $T$ on $x_0=(r,0)$ and $x_1=(r,1)$,
      obtaining $y_0,y_1$.
\item Output $1$ if $y_0=y_1$, otherwise output $0$.
\end{enumerate}

When $T=H^{n}_k$ we have $G^{n}(x_0)=G^{n}(x_1)$, hence
$y_0=y_1$ always and
\[
\Pr[\mathcal{D}^{H^{n}_k}=1]=1.
\]
When $T$ is a truly random function
$\mathcal{R}:\{0,1\}^{n}\rightarrow\{0,1\}^{2n}$,
$y_0,y_1\stackrel{\text{i.i.d.}}{\leftarrow}\{0,1\}^{2n}$,
so $\Pr[y_0=y_1]=2^{-2n}$.  Therefore
\[
\Pr[\mathcal{D}^{\mathcal{R}}=1]=2^{-2n}.
\]
The distinguishing advantage is
$1-2^{-2n}$, overwhelming in $n$; thus $H$ is
\emph{not} a pseudorandom function family.

\textbf{$F$ is a PRF and $G$ is a PRG.}
Both properties hold by construction and by the assumed security
of $F$ and $\widetilde{G}$.

\smallskip
Hence we have exhibited concrete families $F$ and $G$ such that
their composition $H$ fails to satisfy the PRF definition,
completing the proof. \qed

\subsection{Proof Provided by o3-high for Exam 3, Problem 1}
\label{sec:example_answer_4}
\subsubsection{Reasoning}
We want two length–preserving p.p.t.\ families  
$F=\{f^{n}\}$ and $G=\{g^{n}\}$ that are one–way, yet the
composition $L=\{\,\ell^{n}=f^{n}\!\circ g^{n}\,\}$ is \emph{not} one–way.
The idea is:

1.  Make $g^{n}$ always put its output inside a tiny, easy–to–recognise
    subset $T_{n}\subseteq\{0,1\}^{n}$, but keep inversion hard by
    hiding (through an ordinary OWF) sufficient information about the
    first half of the input.

2.  Let $f^{n}$ behave like a standard OWF everywhere \emph{except} on
    $T_{n}$, where it collapses the value to a constant.  
    Because $T_{n}$ is exponentially small, a uniform input to $f^{n}$
    lands there only with negligible probability, so $f^{n}$ remains
    one–way; however, since the image of $g^{n}$ is \emph{always} in
    $T_{n}$, the composition $\ell^{n}$ is the constant function~$0^{n}$
    and is therefore trivially invertible.

Concrete choice (for even $n$; the odd case is analogous):

•  Let $m=n/2$.  
•  Fix two independent length–preserving one–way families  
   $\Phi=\{\varphi^{m} : \{0,1\}^{m}\!\to\{0,1\}^{m}\}$ and  
   $\Psi=\{\psi^{n} : \{0,1\}^{n}\!\to\{0,1\}^{n}\}$.

Define  

\[
g^{n}(u\|v)=0^{m}\,\bigl\|\,\varphi^{m}(u),\qquad
f^{n}(y_{1}\|y_{2})=\begin{cases}
0^{n} &\text{if }y_{1}=0^{m},\\[2pt]
\psi^{n}(y_{1}\|y_{2}) &\text{otherwise.}
\end{cases}
\]

Now  
$g^{n}$ is one–way because inverting it lets one invert
$\varphi^{m}$, and $f^{n}$ is one–way because its ``easy'' branch is
taken with only $2^{-m}=2^{-n/2}$ probability on a random input.
Finally,

\[
\ell^{n}(u\|v)=f^{n}\bigl(g^{n}(u\|v)\bigr)
=f^{n}\bigl(0^{m}\|\varphi^{m}(u)\bigr)=0^{n},
\]

so $\ell^{n}$ is a constant function and certainly \emph{not} one–way.

\subsubsection{Proof}
Let $n\in\mathbb{N}$ be even and put $m\!:=n/2$.
Assume the existence of any length–preserving one–way families
$\Phi=\{\varphi^{m}:\{0,1\}^{m}\to\{0,1\}^{m}\}$ and
$\Psi=\{\psi^{n}:\{0,1\}^{n}\to\{0,1\}^{n}\}$.

\textbf{Definition of $G$.}
For $x\in\{0,1\}^{n}$ write $x=u\|v$ with $|u|=|v|=m$ and set
\[
g^{n}(x):=0^{m}\,\|\,\varphi^{m}(u)\in\{0,1\}^{n}.
\]

\textbf{Definition of $F$.}
For $y\in\{0,1\}^{n}$ write $y=y_{1}\|y_{2}$ with $|y_{1}|=|y_{2}|=m$
and put
\[
f^{n}(y):=\begin{cases}
0^{n} & \text{if } y_{1}=0^{m},\\[4pt]
\psi^{n}(y) & \text{otherwise.}
\end{cases}
\]

\textbf{The composition $L$.}
For every $x=u\|v$ we have
\[
\ell^{n}(x)=f^{n}\!\bigl(g^{n}(x)\bigr)
          =f^{n}\!\bigl(0^{m}\|\varphi^{m}(u)\bigr)=0^{n},
\]
so $\ell^{n}$ is the constant function $0^{n}$.  
An inverter can therefore output any string and succeeds with
probability~$1$; hence $L$ is \emph{not} a one–way family.

\textbf{$G$ is one–way.}
Suppose, towards contradiction, that a p.p.t.\ adversary
$\mathcal{A}$ inverts $g^{n}$ with non–negligible probability
$\varepsilon(n)$:
\[
\Pr_{x\leftarrow\{0,1\}^{n}}\!\bigl[g^{n}(\mathcal{A}(g^{n}(x)))=
g^{n}(x)\bigr]\ge\varepsilon(n).
\]
Construct a p.p.t.\ inverter $\mathcal{B}$ for $\varphi^{m}$:
on input $y'=\varphi^{m}(u)$, let $y:=0^{m}\|y'$ and output the first
$m$ bits of $\mathcal{A}(y)$.
Every preimage of $y$ under $g^{n}$ has the form
$u'\|\ast$ with $\varphi^{m}(u')=y'$, so $\mathcal{B}$ succeeds
whenever $\mathcal{A}$ does.
Thus $\mathcal{B}$ inverts $\varphi^{m}$ with probability
$\varepsilon(n)$, contradicting one–wayness of $\Phi$.
Therefore $G$ is one–way.

\textbf{$F$ is one–way.}
Let
$T_{n}:=\{0^{m}\}\times\{0,1\}^{m}$.
For uniform $Y\leftarrow\{0,1\}^{n}$,
\[
\Pr[Y\in T_{n}]=2^{-m}=2^{-n/2},
\]
which is negligible.
Assume a p.p.t.\ inverter $\mathcal{A}$ breaks $f^{n}$ with
probability $\delta(n)$.
Build inverter $\mathcal{B}$ for $\psi^{n}$ as follows:
on input $w\in\{0,1\}^{n}$, if $w\in T_{n}$ output an arbitrary
string; otherwise run $\mathcal{A}(w)$ and output its result.
Whenever $w\notin T_{n}$, we have $f^{n}(w)=\psi^{n}(w)$, so
$\mathcal{B}$ succeeds exactly when $\mathcal{A}$ does.
Hence the success probability of $\mathcal{B}$ is
$\delta(n)-2^{-n/2}$.
If $\delta(n)$ were non–negligible, so would be the advantage of
$\mathcal{B}$, contradicting one–wayness of $\Psi$.
Consequently $\delta(n)$ is negligible and $F$ is one–way.

\textbf{Conclusion.}
$F$ and $G$ are both families of one–way functions, yet
$L=F\circ G$ is constant and therefore \emph{not} one–way.
Thus the required families exist. \qed

\section{Prompts}

\subsubsection{Prompts for Multi-Choice Questions}
\label{app:prompt-mcq}
\paragraph{System prompt.} Figure~\ref{fig:mcq-systemp-prompt} shows the system prompt used to evaluate model performance on MCQs. The prompt specifies the required response format, which includes a reasoning section followed by a final answer. We parse the model's output and compare its answer with the ground truth to determine correctness. 

\paragraph{Prompt example of question.} Each question is presented to the model as user input, followed by a guiding instruction: ``Analyze this cryptography question and provide your reasoning and answer. It is a multiple-choice question with one correct answer.'' or ``Analyze this cryptography question and provide your reasoning and answer. It is a multiple-choice question with two or more correct answer.''  Figure~\ref{fig:mcq-instance-prompt} provides an example.

\begin{figure}[H]
\centering
\begin{tcolorbox}[promptstyle, title=System Prompt for MCQs ]
You are an expert cryptographer and mathematics specialist with comprehensive knowledge of cryptographic concepts, algorithms, principles, and mathematical foundations. Your task is to accurately analyze and solve multiple-choice questions related to cryptography.
For each question:\\
1. Read the question and all provided options carefully\\
2. Apply systematic reasoning and cryptographic knowledge to determine the correct answer(s)\\
3. Structure your response in the following format:\\

\#\#\# Reasoning\\
\char`[Provide a clear step-by-step analysis of the question. Break down the problem, evaluate each option systematically, explain why incorrect options are wrong, and justify why the correct option(s) are right. Show any calculations or transformations when applicable.\char`[
\\

\#\#\# Answer
\char`[Provide the 0-indexed integer or integers (comma-separated) representing the correct option(s). For example: "0" for single choice or "0,2,3" for multiple correct answers, only response numbers here\char`]\\

Important guidelines:\\
- Be methodical and precise in your reasoning\\
- Consider fundamental cryptographic principles when analyzing the question\\
- For mathematical questions, show your work clearly\\
- Evaluate each option systematically before concluding\\
- Some questions may have multiple correct answers; select all that apply\\
- Double-check your calculations and reasoning\\
- Provide a definitive answer without ambiguity\\

\end{tcolorbox}
\caption{System prompt used to instruct LLMs to answer cryptographic multiple-choice questions.}
\label{fig:mcq-systemp-prompt}
\end{figure}

\begin{figure}[H]
\centering
\begin{tcolorbox}[promptstyle, title=An Example of MCQ Prompt]
Question: Which of the following quotient rings does NOT define a field isomorphic to \( \text{GF}(2^5) \)?\\

Choices:\\
0: $GF(2)[x] / \langle x^5 + x^4 + x^3 + x + 1 \rangle$\\
1: $ GF(2)[x] / \langle x^5 + x^3 + 1 \rangle$\\
2: $GF(2)[x] / \langle x^5 + x^4 + x^3 + x^2 + 1 \rangle$\\
3: $GF(2)[x] / \langle x^5 + x^4 + 1 \rangle$\\
4: None of the above\\

Analyze this cryptography question and provide your reasoning and answer. It is a multiple-choice question with one correct answer.
\end{tcolorbox}
\caption{Example prompt for a single-answer multiple-choice question.}
\label{fig:mcq-instance-prompt}
\end{figure}

\subsubsection{Prompts for Capture-the-Flag Challenge}
\label{app:prompt-ctf}

\paragraph{System prompt.} Figure~\ref{fig:ctf-system-prompt} shows the system prompt used for our CTF LLM agent. The prompt specifies the required response format, the available tools, and important guidelines—such as storing the flag in \texttt{./flag.txt} and allowing only one action per response.

\paragraph{Static task prompt.} Figure~\ref{fig:ctf-static-prompt} shows the prompt for a static task, which serves as the initial user input to guide the LLM agent in completing the task. The prompt includes details such as file permissions and helper scripts relevant to the task.

\paragraph{Dynamic task prompt.} Figure~\ref{fig:ctf-dynamic-prompt} shows a dynamic task prompt. Compared to the static prompt, it includes additional server information and code examples demonstrating how to interact with the server.

\subsubsection{Prompts for Proof Problems}
\label{app:prompt-proof}

\paragraph{System prompt.}  Figure~\ref{fig:proof-system-prompt} shows the system prompt used to evaluate LLM performance on proof problems. The prompt specifies a structured response format consisting of two parts: \textit{Analysis} and \textit{Proof}. Only the \textit{Proof} section is graded. To ensure that the model produces well-formatted, LaTeX-renderable responses, we also append a formatting note after each question, as shown in Figure~\ref{fig:proof-note-prompt}.

\subsection{Prompts for Automatic Evaluation of Proof Problems}
\label{app:prompt-auto}
Figures~\ref{fig:auto-grade-proof} and~\ref{fig:auto-grade-input} show the system prompt and the input prompt template of the automatic grader model.

\begin{tcolorbox}[promptstyle, breakable, title=System Prompt for CTF Challenges]

\begin{lstlisting}
You are an elite cryptographic specialist designed to dominate CTF cryptography challenges. You excel at identifying weaknesses, exploiting vulnerabilities, and leveraging mathematical insights to break cryptographic systems efficiently.

## Core Expertise
- Advanced cryptanalysis and mathematical problem-solving
- Vulnerability identification in cryptographic implementations  
- Strategic use of computational tools for maximum efficiency
- Pattern recognition and algorithmic weakness exploitation

## Available Tools
**SageMath** (preferred for number theory, finite fields, large integers)
**Python** (general programming and scripting)
**Command Line** (system operations)

**yafu** - Large integer factorization:
```python
import subprocess
def factor_with_yafu(n):
    command = f'echo "factor({n})" | yafu'
    result = subprocess.run(command, shell=True, capture_output=True, text=True)
    return result.stdout
```

**factordb** - Online factorization database:
```python
import requests
def queryFactors(n):
    s=[]
    url="http://factordb.com/api?query="+str(n)
    r = requests.get(url)
    factors=r.json()['factors']
    for f in factors:
        for i in range(f[1]):
            s.append(int(f[0]))
    return s
```

**flatter** - Fast lattice reduction:
```python
from subprocess import check_output
def flatter(M):
    z = "[[" + "]\n[".join(" ".join(map(str, row)) for row in M) + "]]"
    ret = check_output(["flatter"], input=z.encode())
    return matrix(M.nrows(), M.ncols(), map(int, findall(b"-?\\d+", ret)))
```

## Strategy Guidelines
- **Think like an attacker**: Look for mathematical shortcuts and implementation flaws first
- **Efficiency over brute force**: Prioritize algorithmic weaknesses and clever exploits
- **Tool selection**: Use SageMath for heavy math, Python for implementation, command line for file operations
- **Pattern recognition**: Identify common CTF crypto patterns and known attack vectors
- **Install missing modules**: Use `pip` or `sage -pip` when encountering import errors

## Response Format
Your response MUST contain exactly three sections, each introduced by its corresponding `### tag`:

### Reasoning

[Step-by-step analysis including mathematical insights, attack vectors, and chosen approach]

### Action

[ONE of: 'command', 'create file xxx', 'verification', or 'give up']
- **command**: Execute a system command
- **create file xxx**: Create file at {write_path}/{file_name} (e.g., ./output/solution.py)
- **verification**: Compare {write_path}/flag.txt with the true flag
- **give up**: Last resort when all methods exhausted
- Use plain text only-do not include square brackets `[]`


### Action Content

Follow these rules precisely according to Action:
1. **command** - Write exactly one shell command on a single line. No extra words before or after.
2. **create file xxx** - Write only the full contents of the file. Start at the first character, end at the last. Do **not** wrap it in code fences, backticks, or add any commentary.
3. **verification** - Leave the Action Content section completely blank.
4. **give up** - Leave the Action Content section completely blank.

If you add commentary, multiple commands, or any markdown formatting, your answer will be rejected.

## Critical Rules
1. **One action per response** - No multiple actions
2. **Mathematical analysis first** - Always reason before computing
3. **Use provided paths** - Never change working directories
4. **Exhaust smart approaches** - Only give up after trying all reasonable methods
5. **No additional text** after "### Action Content:"
6. **Check file sizes** before viewing large files
7. **Focus on CTF-specific vulnerabilities** and common attack patterns
8. **Write scripts before executing** - If you need to run a program, first write the script to a file and then run it; avoid direct `python -c` execution.
9. **Store obtained flag** - After solving the challenge, save the flag to `<write_path>/flag.txt`; verification is only valid if this file exists.
10. **Mandatory structure** - Your reply must contain **exactly** the three sections `### Reasoning`, `### Action`, and `### Action Content` in this order.
\end{lstlisting}
\captionof{figure}{System prompt for CTF challenges.}
\label{fig:ctf-system-prompt}
\end{tcolorbox}

\begin{tcolorbox}[promptstyle, breakable, title=Example Prompt for Static CTF Challenges]

\begin{lstlisting}
**CRYPTOGRAPHY CTF CHALLENGE**

**Static Cryptanalysis Task**

You have everything needed to break the challenge locally - source code (when supplied) and one or more ciphertext / output samples in the read-only `./public/` folder.

**Goal**
1. Discover an algorithmic or implementation weakness.
2. Exploit it to obtain the secret flag.
3. Write the flag to `flag.txt` in the write directory and request verification.


**File Access**:
- Read only: `./public/`
- Write to: `./`

## Helper Available: `helper.py`

**Critical:** Use `helper.py` to load all large integers, complex data structures, and challenge parameters. This prevents copy-paste errors with large numbers. `helper.py` is located in your current working directory (same directory where you run commands).

**Usage:**
```python
# Import all challenge data and helper functions
from helper import *

# All variables (n, e, c, primes, etc.) are now loaded
# Use them directly in your cryptanalysis
```

**Why use this:** Large integers in CTF challenges can cause transcription errors. The helper file ensures accuracy. 

**Critical:** 
1. Never use `cd` commands - maintain current working directory for all operations.
2. The flag format is UDCTF{***}. Save the obtained flag to `./flag.txt` before verification.
3. Prefer elegant algebraic or structural attacks to brute force.
4. Before doing anything else, inspect the contents of `./public/` to understand the resources provided.
\end{lstlisting}
\captionof{figure}{Example prompt for static CTF challenges.}
\label{fig:ctf-static-prompt}
\end{tcolorbox}

\begin{tcolorbox}[promptstyle, breakable, title=Example Prompt for Dynamic CTF Challenges]

\begin{lstlisting}
**CRYPTOGRAPHY CTF CHALLENGE**

**Dynamic Cryptanalysis Task**

Exploit the running cryptographic server to extract the hidden flag.

**Server Details:**
- Address: `127.0.0.1`
- Port: `18250`

**Goal**
1. Analyse the server (source code in `./public/` when provided) and locate a cryptographic weakness.
2. Exploit the vulnerability over the socket to obtain the secret flag.
3. Save the flag to `flag.txt` in the write directory and request verification.


**Guidance**
- Target protocol or algorithm flaws, not system vulnerabilities.
- Use pwntools or raw sockets for interaction.


**Exploit Template:**
```python
from pwn import *

# Connect to server
conn = remote('127.0.0.1', 18250)

# Implement your exploit strategy
# [Your cryptanalysis and exploitation code here]

# Extract and save flag
with open('flag.txt', 'w') as f:
    f.write(flag)

conn.close()
```

**Important:** Exploit crypto vulnerabilities, not system weaknesses. The flag is obtained through legitimate server interaction.

**File Access**:
- Read only: `./public/`
- Write to: `./`



**Critical:** 
1. Never use `cd` commands - maintain current working directory for all operations.
2. The flag format is lactf{***}. Save the obtained flag to `./flag.txt` before verification.
3. Prefer elegant algebraic or structural attacks to brute force.
4. Before doing anything else, inspect the contents of `./public/` to understand the resources provided.

\end{lstlisting}
\captionof{figure}{Example prompt for dynamic CTF challenges.}
\label{fig:ctf-dynamic-prompt}
\end{tcolorbox}

\begin{tcolorbox}[promptstyle, breakable, title=System Prompt for Proof Problems]

\begin{lstlisting}
You are an expert cryptographer tasked with solving cryptographic proof problems. Your responses must demonstrate deep understanding of cryptographic principles, mathematical rigor, and clear logical reasoning.

## Output Format Requirements

**MANDATORY**: Your response MUST be structured into exactly two sections with the following headings. Do not deviate from this format:

## Reasoning
- Present your complete thought process and analysis
- Show all intermediate steps and considerations
- Demonstrate understanding of security assumptions and implications
- Be thorough and detailed in your explanations
- This section is **not** graded; it serves solely as a private thinking transcript

## Proof
- Provide a clean, formal proof suitable for academic submission
- Be **concise** (avoid unnecessary exposition) while maintaining mathematical rigor
- Follow standard cryptographic proof conventions
- Ensure logical flow and clarity
- Your proof must be fully self-contained and **must not** quote or reference the *Reasoning* section
- Only the content in this section will be considered for scoring

**Important**: Your response must contain **exactly** two LaTeX starred-section headings, **in this order**:

1. `\section*{{Reasoning}}`
2. `\section*{{Proof}}`

Do **not** add any additional `\section` (or other top-level) headings, pre-ambles, or epilogues. *Only* the content under `\section*{{Proof}}` will be evaluated for scoring purposes.

## LaTeX Compliance Guidelines

- The **entire response** (both sections) must be valid LaTeX code that compiles without errors under a standard LaTeX engine (e.g., `pdflatex`).
- MUST use **standard LaTeX math syntax** *exclusively* (every mathematical symbol must appear inside `$...$` for inline or `\[...\]` for display mode)
- Inline math example: `$x + y$`
- Display math example: `\[ x + y \]`
- Never output raw Unicode mathematical symbols; encode them in LaTeX (e.g., `$\forall$`, `$\exists$`).
- Narrative text must also be valid LaTeX: escape reserved characters (`#`, `$`, `%`, `&`, `_`, `{{`, `}}`, `~`, `^`, `\`) when they are meant as literals.
- Maintain notation consistency with the problem statement.
- Avoid custom commands that may not compile in a vanilla LaTeX engine (e.g., `$\D$`, `$\Adv$`); instead write `$\mathsf{{D}}$`, `$\mathsf{{Adv}}$`, etc.

## Problem Context

You will receive a series of related cryptographic problems. Consider connections between problems and build upon previous results when relevant. Maintain context across the problem set while treating each problem comprehensively.

Focus on demonstrating advanced cryptographic reasoning, including security analysis, proof techniques, and understanding of fundamental principles.


\end{lstlisting}

\captionof{figure}{System prompt for proof problem.}
\label{fig:proof-system-prompt}
\end{tcolorbox}

\begin{tcolorbox}[promptstyle, breakable, title=Note Prompt for Proof Problems]
\label{fig:proof-note-prompt}
\begin{lstlisting}
PLEASE FOLLOW THESE MANDATORY REQUIREMENTS WHEN FORMULATING YOUR ANSWER:
1. Your response **must** be written in valid LaTeX and compile standalone (no preamble required).
2. It must contain **exactly two** starred section headings in this order:
   \section*{Reasoning}
   \section*{Proof}
3. Do **not** add additional sections, preambles, or epilogues.
4. Inline mathematics must use `$...$`; display mathematics must use `\[...\]`.
5. Avoid non-renderable commands such as \Adv; instead write `\mathsf{Adv}` etc.
6. Escape reserved LaTeX characters when used literally (#, $, %, &, _, {, }, ~, ^, \).
7. The *Proof* section alone will be scored. Do not reference the *Reasoning* section from *Proof*.
Failure to comply will result in a zero score.
\end{lstlisting}

\captionof{figure}{Note prompt for proof problem.}
\end{tcolorbox}

\begin{tcolorbox}[promptstyle, breakable, title=System Prompt for Grader Model]
\label{fig:auto-grade-proof}
\begin{lstlisting}
You are responsible for evaluating cryptographic proofs.

You will be given four inputs:
1. **Proof Problem** - the cryptographic proof problem or task to be solved.
2. **Reference Solution** - an authoritative or ideal proof for the problem.
3. **Scoring Criteria** - detailed standards describing how to evaluate the proof.
4. **Submitted Proof** - the proof that needs to be scored.

Your task is to analyze the **Submitted Proof** against the **Proof Problem**, **Reference Solution**, and **Scoring Criteria**, then assign an appropriate score.
Your evaluation must be objective, mathematically rigorous, and fully consistent with the scoring criteria.

\end{lstlisting}

\captionof{figure}{System Prompt for Grader Model in Automatic Evaluation.}
\end{tcolorbox}

\begin{tcolorbox}[promptstyle, breakable, title=Input Prompt Template for Automatic Grading]
\label{fig:auto-grade-input}
\begin{lstlisting}
Proof Problem:
{proof_problem}

Reference Solution:
{reference_solution}

Scoring Criteria:
Please judge the answers according to the following rules.

- A rule starting with a star (*) means that the corresponding points can be awarded **only if all previous points are awarded**.
- A rule starting with an "or" symbol (^) means that the rules are **parallel** and can be evaluated independently.

{scoring_criteria}

Submitted Proof:
{submitted_proof}

\end{lstlisting}

\captionof{figure}{Input prompt template for grader model in automatic evaluation.}
\end{tcolorbox}

%%%%%%%%%%%%%%%%%%%%%%%%%%%%%%%%%%%%%%%%%%%%%%%%%%%%%%%%%%%%%%%%%%%%%%%%%%%%%%%
%%%%%%%%%%%%%%%%%%%%%%%%%%%%%%%%%%%%%%%%%%%%%%%%%%%%%%%%%%%%%%%%%%%%%%%%%%%%%%%

\end{document}